\documentclass[twocolumns,a4paper]{aa}
\usepackage{graphicx}
\usepackage{txfonts}
\usepackage{color}
\usepackage{natbib}
\usepackage{hyperref}

\usepackage{amsmath}

\definecolor{blue}{RGB}{0,0,255}
\definecolor{red}{RGB}{255,0,0}
\definecolor{green}{RGB}{0,255,0}

\newcommand{\helixp}{\textsc{HeLIx$^+$}}

\begin{document}

\title{Combining magneto-hydrostatic constraints with Stokes profiles inversions.\\
III. Uncertainty in the inference of electric currents}

\author{J.M.~Borrero\inst{1} \and A.~Pastor Yabar\inst{2}}

\institute{Leibniz-Institut f\"ur Sonnenphysik, Sch\"oneckstr. 6, D-79104, Freiburg, Germany
\and
Institute for Solar Physics, Department of Astronomy, Stockholm University, AlbaNova University 
Centre, 10691 Stockholm, Sweden}
\date{Recieved / Accepted}

\abstract{Electric currents play an important role in the energy balance of the plasma in the solar atmosphere. 
They are also indicative of non-potential magnetic fields and magnetic reconnection. Unfortunately, the direct 
measuring of electric currents has traditionally been riddled with inaccuracies.}{We study how accurately we can infer 
electric currents under different scenarios.}{We carry out increasingly complex inversions of the 
radiative transfer equation for polarized light applied to Stokes profiles synthesized from radiative three-dimensional 
magnetohydrodynamic (MHD) simulations. The inversion yields the magnetic field vector, ${\bf B}$, from which the electric current density,
${\bf j}$, is derived by applying Ampere's law.}{We find that the retrieval of the electric current density is only slightly affected
by photon noise or spectral resolution. However, the retrieval steadily improves as the Stokes inversion becomes increasingly 
elaborated. In the least complex case (a Milne-Eddington-like inversion applied to a single spectral region), it 
is possible to determine the individual components of the electric current density ($j_{\rm x}$, $j_{\rm y}$, $j_{\rm z}$)
with an accuracy of $\sigma=0.90-1.00$ dex, whereas the modulus ($\|{\bf j}\|$) can only be 
determined with $\sigma=0.75$ dex. In the most complicated case (with multiple spectral regions, a large number of nodes, Tikhonov vertical regularization, and magnetohydrostatic equilibrium), these numbers improve to $\sigma=0.70-0.75$
dex for the individual components and $\sigma=0.5$ dex for the modulus. Moreover, in regions where the magnetic field is above 300 gauss,
$\|{\bf j}\|$ can be inferred with an accuracy of $\sigma=0.3$ dex. In general, the $x$ and $y$ components of the electric current density  are retrieved slightly better than the $z$ component. In addition, the modulus of the electric current density is the best
retrieved parameter of all, and thus it can potentially be used to detect regions of enhanced Joule heating.}
{The fact that the accuracy does not worsen with decreasing spectral resolution or increasing photon noise, and instead
increases as the Stokes inversion complexity grows, suggests that the main source of errors in the determination of 
electric currents is the lack of realism in the inversion model employed
to determine variations in the magnetic field along the line of sight at scales smaller than the photon mean-free path, along with the intrinsic limitations of the model due to radiative transfer effects.}

\titlerunning{Accuracy in the Inference of electric currents on the Sun}
\authorrunning{Borrero \& Pastor Yabar}
\maketitle

\def\kms{~km s$^{-1}$}
\def\deg{^{\circ}}
\def\df{{\rm d}}
\newcommand{\ve}[1]{{\rm\bf {#1}}}
\newcommand{\diff}{{\rm d}}

\section{Introduction}
\label{sec:introduction}

Electric currents, ${\bf j}$, in the solar atmosphere play a very important role. They are crucial
for the energy balance through Joule and ambipolar heating \citep{priest1999heating,cheung2012,lena2021ambipolarhall}.
They can also be employed as proxies of regions where the magnetic field is highly non-potential
and therefore likely to cause transient and explosive events in the solar chromosphere and
corona \citep{priest2002flare,green2018review}. Consequently, measuring electric currents
has been long considered one of the most important goals in solar physics \citep{jan2019est}.
The inference of electric currents is possible via magnetic field extrapolations \citep{rohan2021j}. 
However, the extrapolated magnetic field differs both in magnitude and fine structure from the actual 
measurements \citep{vissers2022}, thereby casting doubts on the electric currents inferred in this fashion.\\ 

It is therefore more compelling to use the measured magnetic field instead. Most previous determinations 
of electric currents based on the measured ${\bf B}$ were carried out through Stokes inversion techniques that employ
the Milne-Eddington (ME) approximation \citep{solanki2003he,wang2017flare}. However, ME inversion techniques have been used to
calculate only the vertical component of the electric current density, $j_z$, instead of the full
vector, ${\bf j}$. This occurs because ME inversion yields constant values along the $z$ direction \citep{auer1977, landolfi1982} 
and therefore lacks the information needed to fully evaluate $j_x$ and $j_y$. Non-ME Stokes 
inversion codes are indeed capable of inferring ${\bf j}$, but they do so by calculating a $z$ scale that is based on 
hydrostatic equilibrium \citep[HE;][]{hector2005currents}. To our knowledge, the only work in which the three components of 
the electric current density, ${\bf j}$, were determined from the measured magnetic field ${\bf B}$, not using HE, 
is \citet{puschmann2010currents}. This work is instead based on the assumption of magnetohydrostatic (MHS)
equilibrium \citep[][]{puschmann2010pen}.\\

Despite being of great importance, the study by \citet{puschmann2010pen,puschmann2010currents} suffers from a number of shortcomings
\citep{borrero2019mhs}. In \citet{borrero2021mhs} we present a novel Stokes inversion code that also employs MHS 
equilibrium and solves most of those shortcomings. The resulting method was subsequently employed to determine electric
currents in the solar photosphere and compared to results from magnetohydrodynamic (MHD)  simulations \citep{adur2021currents}, thereby allowing
us to study,  for the first time, the accuracy in the inference of ${\bf j}$. However, a missing piece
of the puzzle is whether this accuracy actually improves results that employ MHS instead of HE and, if so,
by how much. In addition, we study the effects of, among other factors, spectral resolution, photon noise, and the number
of spectral lines inverted. To this end, in this work we again employ MHD simulations to calculate simulated
Stokes profiles in a number of spectral lines (Sect.~\ref{sec:synthesis}). These Stokes profiles are then subjected
to a number of Stokes inversion techniques with increasing degrees of complexity (including HE, MHS equilibrium, etc;
Sect.~\ref{sec:inference}). Results are discussed and compared in Sect.~\ref{sec:discussion}. The effects
of photon noise and limited spectral resolution are studied in Sect.~\ref{sec:noise_spec}. Finally, Sect.~\ref{sec:conclu}
summarizes our findings.\\

\section{MHD simulations and synthetic Stokes profiles}
\label{sec:synthesis}

\begin{figure*}
\centering
   \includegraphics[width=18cm]{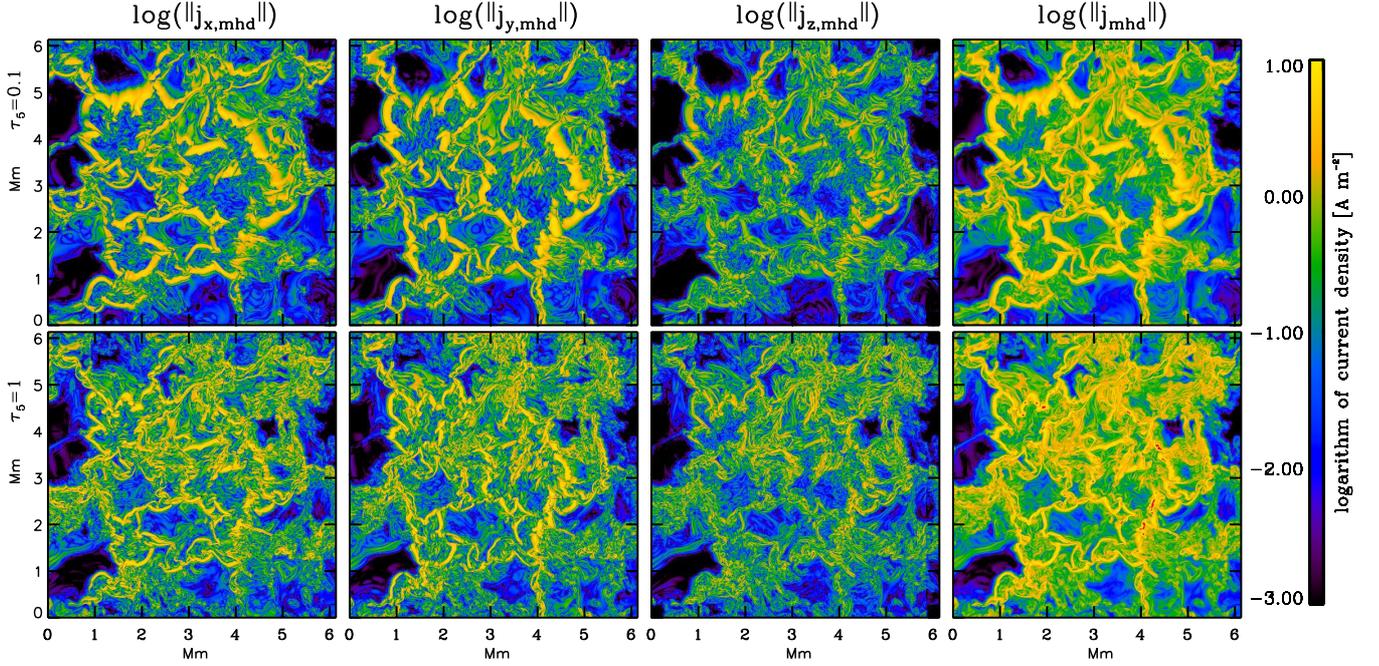}
   \caption{Logarithm of the absolute value of the electric current density in the MHD simulations. {\it From left to 
right}: $j_{\rm x,mhd}(x,y,\tau_c)$, $j_{\rm y,mhd}(x,y,\tau_c)$, $j_{\rm z,mhd}(x,y,\tau_c)$, and $\|\ve{j}_{\rm mhd}(x,y,\tau_c)\|$. 
The electric currents are shown at two different optical depths: $\tau_c=0.1$ (upper panels) and $\tau_c=1$ (lower panels).
\label{fig:currents2d_mhd}}
\end{figure*}

In this paper we employ the same physical parameters as in \citet{adur2021currents}. These parameters come from one 
single snapshot that resulted from the three-dimensional radiative MHD simulations by \citet{rempel2012mhd}. 
The original simulation box has a size of 4096$\times$4096$\times$768 cells, with the first two values corresponding 
to the solar surface, $XY$, and the third one, $Z$,  perpendicular to the surface (i.e., antiparallel to the Sun's gravity 
direction). The grid spacing is $\Delta x \times \Delta y \times \Delta z=12\times 12\times 8$ km in each spatial direction. 
The simulation setup is that of a large sunspot surrounded by granulation and different magnetic knots. From here we 
selected a smaller region of 512$\times$512 cells on the $XY$ plane. This region includes a couple of pores embedded  
in granulation \citep[see Appendix B in][]{adur2021currents}. The magnetic field from the simulations, $\ve{B}_{\rm mhd}$, was 
employed to determine the electric current density by applying Ampere's law: $\ve{j} = \mu_0^{-1} \nabla \times {\bf B}$, 
where $\ve{j}$ is given in A m$^{-2}$ when $\ve{B}$ is given in teslas, and $\mu_0= 4 \pi 10^{-7}$ kg m s$^{-2}$ A$^{-2}$ is 
the vacuum magnetic permeability. Figure~\ref{fig:currents2d_mhd} illustrates the logarithm of the absolute value of the 
three components of electric current density, $\ve{j}_{\rm mhd}$: $j_{\rm x,mhd}$ (first column), $j_{\rm y,mhd}$ (second column), and 
$j_{\rm z,mhd}$ (third column), as well as its modulus, $\|\ve{j}_{\rm mhd}\|=\sqrt{j_{\rm x}^2+j_{\rm y}^2+j_{\rm z}^2}$ (rightmost column),
in SI units of amperes per m$^{2}$.\\

\begin{table*}
\caption{Spectral lines and their associated atomic parameters for the three spectral ranges considered.}
\label{tab:spectral_ranges}
\centering                 
\begin{tabular}{cccccccccc}
\hline
\hline                     
Spectral range & $\Delta\lambda$ & ${\rm n}_{\lambda}$ & $\lambda_{0}$ & ${\rm e}_{\rm low}$ & ${\rm e}_{\rm upp}$ & 
$\log_{10}{{\rm g}\,{\rm f}}$ & ${\rm E}_{\rm low}$ & $\alpha$ & $\sigma/{\rm a}_{0}^{2}$ \\
 $\textrm{[nm]}$ & [m{\AA}] &  & [{\AA}] &  &  &  & [eV] &  &  \\
\hline                     
    630.2 & 20 & 55 & Fe {\sc I} 6301.501 & $^{5.0}{\rm P}_{2.0}$ & $^{5.0}{\rm D}_{2.0}$ & -0.718 & 3.654 & 0.243 & 835.356\\
          & 20 & 55 & Fe {\sc I} 6302.494 & $^{5.0}{\rm P}_{1.0}$ & $^{5.0}{\rm D}_{0.0}$ & -1.236 & 3.686 & 0.239 & 850.930\\
    1082.7 & 18    & 275 & Si {\sc I} 10827.091 & $^{3.0}{\rm P}_{2.0}$ & $^{3.0}{\rm P}_{2.0}$ & 0.239 & 4.953 & 0.231 & 729.807\\
    1565.0 & 40    & 125 & Fe {\sc I} 15648.515 & $^{7.0}{\rm D}_{1.0}$ & $^{7.0}{\rm D}_{1.0}$ & -0.669 & 5.426 & 0.229 & 974.195\\
           & 40    & 125 & Fe {\sc I} 15662.018 & $^{5.0}{\rm F}_{5.0}$ & $^{5.0}{\rm F}_{4.0}$ & 0.19 & 5.83 & 0.240 & 1196.950\\
\hline                                           
\end{tabular}
\tablefoot{The first column provides the label used for each spectral region (which might include more than one spectral line). Regarding the symbols: 
$\Delta\lambda$ is the pixel size in m{\AA}; ${\rm n}_{\lambda}$ is the number of wavelengths used for that spectral line; 
$\lambda_{0}$ is the central wavelength for the electronic transition associated with the spectral line; ${\rm e}_{\rm low}$ and 
${\rm e}_{\rm upp}$ are the electronic configurations of the lower and upper energy level, respectively; and ${\rm E}_{\rm low}$ is 
the excitation potential (in eV) of the lower energy level. These atomic data are adopted from \cite{nave1994}. The
$\alpha$ and $\sigma/{\rm a}_{0}^{2}$ are the velocity exponent and collision cross-section parameters (in units
of Bohr's radius, $a_0$), respectively, as defined in the Anstee, Barklem, and O'Mara collision theory for the broadening of 
metallic lines by neutral hydrogen collisions \citep{anstee1995, barklem1997, barklem1998}.}
\end{table*}

The physical parameters from the MHD simulations were employed to solve the radiative transfer equation for polarized light, using the forward 
module of the FIRTEZ code \citep{adur2019invz} so as to obtain the Stokes vector as a function of wavelength, 
${\varmathbb I}^{\rm syn}(x,y,\lambda)$, in a number of spectral lines. We refer to these as ``synthetic'' observations. Although FIRTEZ's solver for
the radiative transfer equation considers the physical parameters to be constant within each grid cell, in reality they are expected to vary internally. 
The effect of sub-grid variations (i.e., at scales smaller than $\Delta z < 8$~km) is not investigated in this paper.\\

The selected spectral lines, number of wavelengths, spectral sampling\footnote{``{Spectral sampling''} should not be confused with ``{spectral resolution,''} which, 
unless a spectral transmission function is used to convolve the Stokes profiles (see Sect.~\ref{sec:noise_spec}), is in effect infinity.}, 
and atomic parameters are given in Table~\ref{tab:spectral_ranges}. Unless otherwise specified, normally distributed random noise -- with a $\sigma_{pn}=10^{-3}$
(in units of the continuum intensity), where {\it pn} refers to {photon noise} -- was added to ${\varmathbb I}^{\rm syn}(x,y,\lambda)$ 
to mimic the effects of photon noise. The synthetic observations are not spatially convolved with a point spread function, so the 
equivalent horizontal spatial resolution of the observations is that of the original MHD simulation: 
$\Delta x = \Delta y =12$ km. This is similar to the resolution that new-generation solar telescopes will achieve
\citep{collados2013est,elmore2014dkist,jan2019est,rimmele2020dkist}.\\

\section{Inference of electric currents via Stokes inversion}
\label{sec:inference}

In this section we perform a series of inversions of the {synthetic} observations, ${\varmathbb I}^{\rm syn}(\lambda)$,
described in Sect.~\ref{sec:synthesis} to recover the physical parameters of the solar atmosphere: $T_{\rm inv}(x,y,z)$, 
$\rho_{\rm inv}(x,y,z)$, $P_{\rm g,inv}(x,y,z)$, $\ve{B}_{\rm inv}(x,y,z)$, and $v_{\rm z,inv}(x,y,z)$. The inversions were all 
performed using the FIRTEZ-dz code \citep{adur2019invz}. We note that some other well-known Stokes inversion codes
are able to carry out some of the inversions described in this section. We will mention which codes can be used and when.  
After the inversion, the horizontal components of the magnetic field $(B_x,B_y)$ were corrected to remove the well-known 
180$^{\circ}$ ambiguity by applying the non-potential magnetic field calculation 
\citep[][]{manolis2005}. After this, Ampere's law was used to determine the electric current density, $\ve{j}_{\rm inv}$, 
which was then compared to the currents from the numerical simulations, $\ve{j}_{\rm mhd}$ (see Fig.~\ref{fig:currents2d_mhd}). 
In the following sections (Sects.~\ref{subsec:me}-\ref{subsec:mhs}), Stokes inversions are carried out with increasing 
degrees of complexity to study if and how the inferred electric current density improves.\\

\subsection{Milne-Eddington-like inversion (\emph{me})}
\label{subsec:me}

As a first step we performed an inversion of only the first spectral range in Table~\ref{tab:spectral_ranges}: the 
Fe {\sc I} line pair at 630.2 nm that was typically observed by the Advanced Stokes Polarimeter \citep[ASP;][]{elmore1992asp} 
and later became the lines of choice of the spectropolarimeter on board Hinode \citep{lites2001hinode,ichimoto2007hinode}.\\

The inversion here was carried out along the $z$ direction independently for each $(x,y)$ column with the following numbers of 
nodes: 32 for $T(z)$, 16 for $v_{\rm z}(z)$, 1 for $B_{\rm x}(z)$, 1 for $B_{\rm y}(z)$, and 1 for $B_{\rm z}(z)$. Because only one
node is given to the each of the three components of the magnetic field, FIRTEZ-dz retrieves values that are constant in $z$,
and therefore these results are comparable to a traditional ME inversion, such as those
carried out with the ASP/HAO code \citep{lites1990,lites1993}, \helixp{} \cite[]{lagg2004,lagg2009}, or VFISV \citep{borrero2011vfisv}.
A summary and comparison of these three codes can be found in \citet{borrero2014milne}. However, unlike with ME Stokes inversion codes, 
we still allow for a $z$-dependent $v_{\rm z}$ and $T_{\rm z}$. The temperature in particular is important as 
it means that we solve the Saha and Boltzmann equations (i.e., we assume local thermodynamic equilibrium) instead of parameterizing 
the source function with two free parameters, as done by ME inversion codes. In order to do so, the gas pressure and density must 
also be known. This was accomplished by assuming HE.\\

During the inversion, different weights were given to each of the four Stokes parameters: $w_i=1$, $w_q=w_u=4$, and $w_v=2$. 
We refer to this inversion as \emph{me}.\\

\subsection{Inversion under hydrostatic equilibrium (\emph{he1})}
\label{subsec:hydeq}

The next step was to drop the assumption made by ME codes and allow for the three components of the magnetic field
to vary with height. In this case, the number of nodes was increased to: four for $B_{\rm x}(z)$, four for $B_{\rm y}(z)$, and eight for 
$B_{\rm z}(z)$. The $T(z)$ and $v_{\rm z}$ were again given 32 and 16 nodes, respectively. All other inversion parameters (weights, 
HE, etc.) were the same as in Sect.~\ref{subsec:me}. We refer to the inversion described in this section 
as \emph{he1}. Examples of Stokes inversion codes that are capable of performing this kind of analysis are: SIR \citep{basilio1992sir}, 
SPINOR \citep{frutiger1999spinor,vannoort2012decon}, NICOLE \citep{socas2015nicole}, and SNAPI \citep{milic2018snapi}.\\

\subsection{Increased number of spectral windows (\emph{he2})}
\label{subsec:lines}

In the \emph{he2} case, the Stokes inversion was carried out in the same fashion as for \emph{he1} (HE; 
Sect.~\ref{subsec:hydeq}) but with a greater number of spectral regions, and thus a  greater number of spectral lines, such that all lines in 
Table~\ref{tab:spectral_ranges} were included in the inversion. The idea behind this is twofold: ({\bf a}) the addition
of redundant information about the magnetic field in the region where the Fe {\sc I} line pair at 630.2 nm is formed
should improve the determination of the derivatives of the magnetic field components with respect to the vertical axis, $z$, in
this region, and ({\bf b}) the addition of spectral lines that convey information about the magnetic field outside the region 
where the Fe {\sc I} line pair at 630.2 nm is formed should allow us to determine the electric current density over a wider 
range of heights (or optical depths). All inversion codes mentioned in Sect.~\ref{subsec:hydeq} (\emph{he1}) can also 
perform the inversions described here.\\

\subsection{Increased number of nodes (\emph{he3})}
\label{subsec:nodes}

In the \emph{he3} case, the Stokes inversion was carried out in the same fashion as in \emph{he2} (HE
plus multiple spectral ranges or lines; Sect.~\ref{subsec:lines}) but now with the number of nodes increased to 16 for $B_{\rm x}(z)$,
16 for $B_{\rm y}(z)$, and 32 for $B_{\rm z}(z)$. The idea behind this inversion is that in regions where the kinetic energy is much higher
than the magnetic energy, the MHD simulations feature very large magnetic field variations along the $z$ direction. This occurs
because, in this case, the convective motions in the solar atmosphere braid the magnetic field lines into non-potential configurations
where the magnetic field can present very large variations over small spatial scales. These variations lead, in general, to very
asymmetric Stokes parameters \citep{egidio1996ncp,borrero2007ncp} that can only be reproduced by inversion if enough free parameters are 
given to the three components of the magnetic field. Examples of such large $z$ variations, and how Stokes inversion 
techniques perform with a low number of nodes, can be found in the literature (see, e.g., Fig.~2 in \citealt{luis2006review} and 
Fig.~5 in \citealt{borrero2014milne}). Of course, it is always a wise policy to increase the number of free parameters only if 
enough information is contained in the synthesized spectral lines, and therefore this new degree of complexity was only added 
 in the \emph{he2} case and not \emph{he1} because the former considers more spectral lines. Again, all inversion codes mentioned in 
Sect.~\ref{subsec:hydeq} (\emph{he1}) can also perform the inversions described here.\\
\subsection{Effect of vertical Tikhonov  regularization (\emph{he4})}
\label{subsec:regul}

In the \emph{he4} case, the Stokes inversion was carried out in the same way as in \emph{he3} (HE,
plus multiple spectral ranges or lines, plus an increased number of nodes; Sect.~\ref{subsec:nodes}), but with a Tikhonov-like
regularization included along the $z$ direction. This regularization penalizes solutions where the components of the magnetic field
present very large variations along the $z$ direction, unless those variations are truly needed in order to properly fit the
synthetic Stokes profiles, ${\varmathbb I}^{\rm syn}(x,y,\lambda)$. This was done via a modification of the $\chi^2$ merit function
that is being minimized by the Stokes inversion code \citep{jaime2019}. One might wonder whether limiting the possible vertical 
variations in the three components of the magnetic field works against the previous Stokes inversion (\emph{he3}), where 
we allowed for larger variations along the $z$-axis, and whether, because of this, \emph{he4} is comparable to \emph{he2}. 
This is in principle a valid concern; however, we need to take into consideration the fact that inversion \emph{he2} limited the vertical variations 
everywhere, whereas \emph{he3} allowed for larger vertical variations everywhere. As we pointed out in Sect.~\ref{subsec:nodes}, 
the large numbers of free parameters might be needed only in regions where the kinetic energy is higher 
than the magnetic energy. Therefore, in regions where the opposite occurs (i.e., umbra, pores, etc.), this might not be necessary and could even be 
counterproductive. To this end, the regularization included in \emph{he4} allows the inversion code FIRTEZ-dz 
to choose between either of the aforementioned situations based on whether the fit to the synthetic Stokes profiles truly
improves when large $z$ variations are included; otherwise, it favors smoother solutions along the $z$ direction. Other codes capable of carrying out Stokes inversion with vertical regularization are STiC \citep{jaime2019} and TIC \citep{li2022inv}.\\

\subsection{Inversion under magnetohydrostatic equilibrium (\emph{mhs})}
\label{subsec:mhs}

In this new inversion, which we call \emph{mhs}, we included all previous improvements (multiple spectral lines or ranges), increased number
of nodes, and vertical $z$ Tikhonov regularization), and, in addition, we dropped the assumption of HE and
 instead used MHS equilibrium. This was done following the iterative approach described in 
\citet[][Fig.~2]{borrero2021mhs} but including all terms of the divergence of the Lorentz force (see Eq.~B.7 in that paper). This is the
most complex inversion tested in this paper. Only the FIRTEZ-dz Stokes inversion code is capable of including MHS constraints.\\ 
%
\section{Discussion}
\label{sec:discussion}

\subsection{Summary of results}
\label{subsec:summary}

\begin{figure*}
\centering
   \includegraphics[width=18cm]{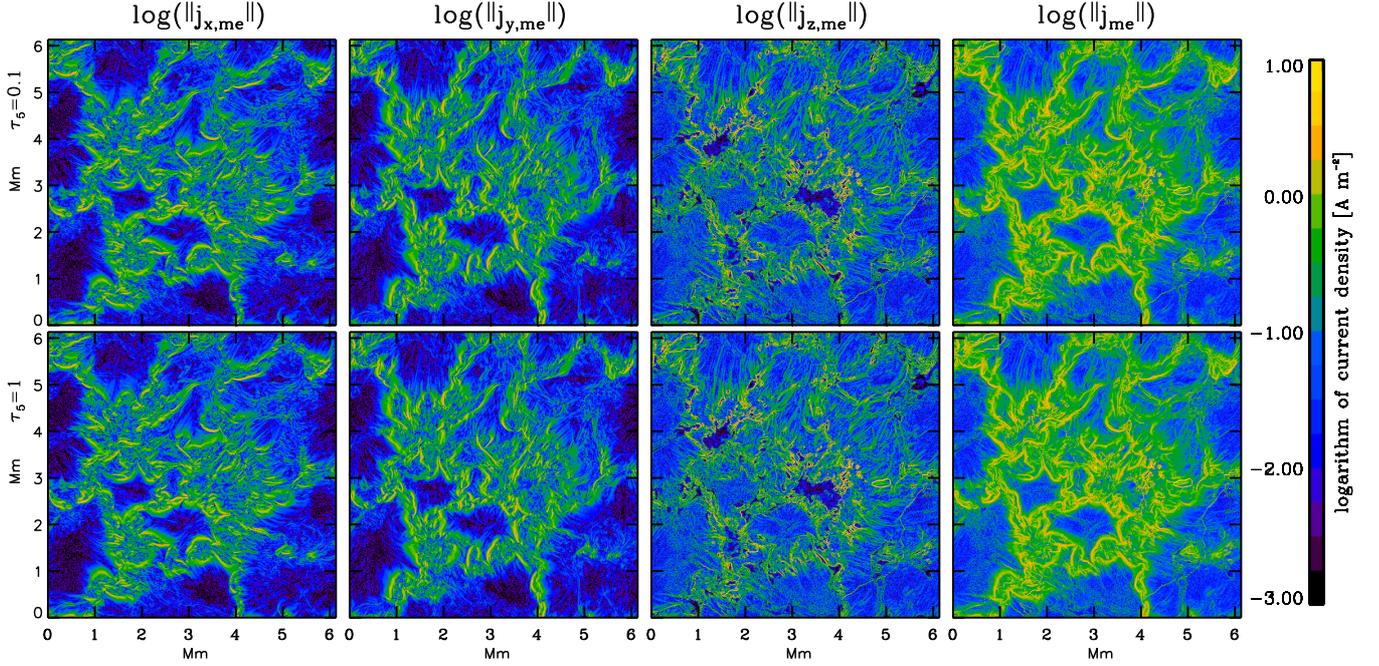}
   \caption{Same as Fig.~\ref{fig:currents2d_mhd} but showing the electric currents as obtained from the ME-like
inversions described in Sect.~\ref{subsec:me} (\emph{he1}).\label{fig:currents2d_me}} 
\end{figure*}
\begin{figure*}
\centering
   \includegraphics[width=18cm]{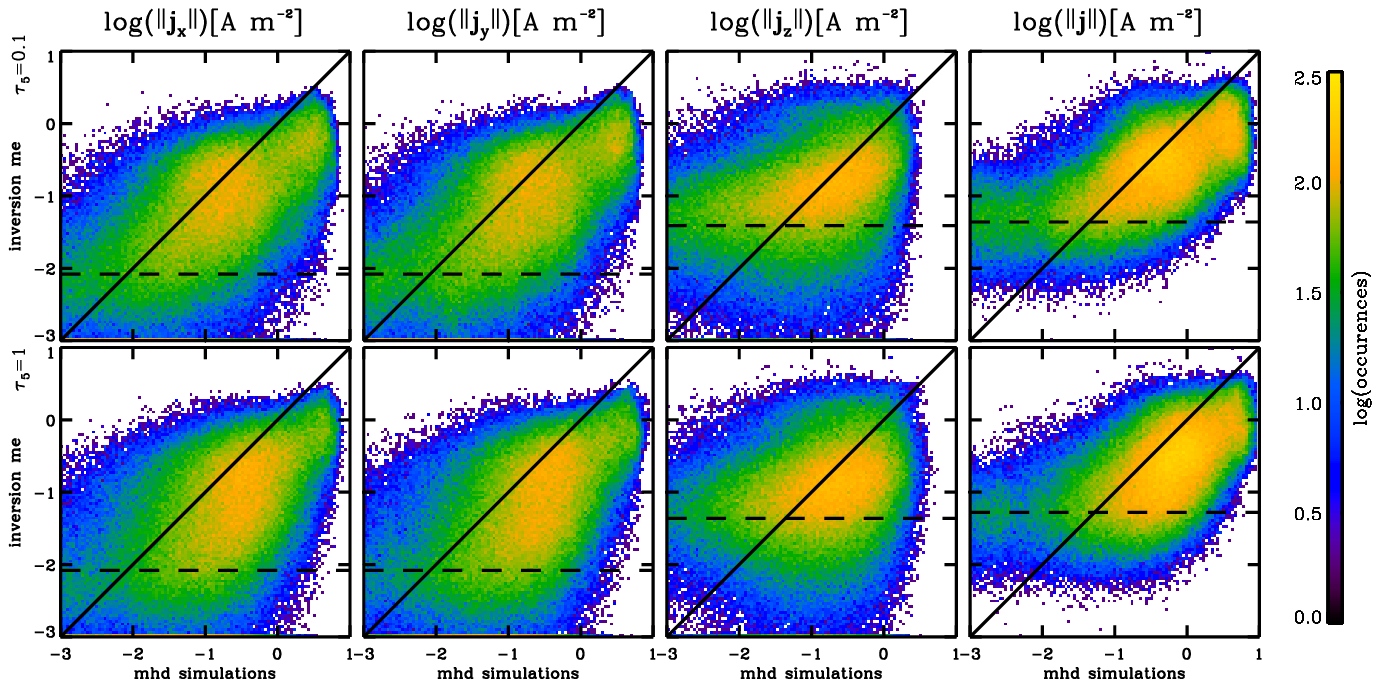}
   \caption{Logarithmic scatter-density plots of the electric current density from the MHD simulations (horizontal axis) and the values inferred
by the \emph{me} Stokes inversion (vertical axis) at $\tau_c=1$ (lower panels) and $\tau_c=0.1$ (upper panels). {\it From left to right}:
Logarithm of the absolute value of the $\|j_{\rm x}\|$ component of the electric current density, logarithm of the absolute value of the $\|j_{\rm y}\|$ 
component of the electric current density, logarithm of the absolute value of the $\|j_{\rm z}\|$ component of the electric current density,
and logarithm of the absolute value of the modulus of the electric current density, $\|\ve{j}\|$. The horizontal dashed black line represents
the so-called {floor} value, below which the Stokes inversion is not capable of correctly inferring the electric currents (see text for
details). The solid thick diagonal corresponds to the one-to-one retrieval along which the Stokes inversion perfectly retrieves the electric
currents. All pixels in the selected region described in Sect.~\ref{sec:synthesis} are included in this figure.\label{fig:currents_me_vs_mhd}}  
\end{figure*}

Let us first briefly consider Fig.~\ref{fig:currents2d_me}. This figure shows the same plot as 
Fig.~\ref{fig:currents2d_mhd} but as inferred from the \emph{me} inversion (Sect.~\ref{subsec:me}). Visual comparison between these 
two figures reveals that the Stokes inversion is in general capable of grasping the overall spatial distribution of the 
electric current density, although the magnitude and horizontal thickness of the current sheets are clearly underestimated by the \emph{me} inversion.
We note that Fig.~\ref{fig:currents2d_me} also includes $j_x$ and $j_y$ despite the ME-like inversion retrieving
$\partial B_y / \partial z = \partial B_x / \partial z =  0$ (see Eq.~\ref{eq:j}). A more detailed comparison can be made from Fig.~\ref{fig:currents_me_vs_mhd}. This new figure presents, from left to right, density plots of the logarithm of the absolute value of $j_{\rm x}$, $j_{\rm y}$, 
$j_{\rm z}$, and $\|\ve{j}\|$ in the MHD simulations (horizontal axis) versus those inferred from the \emph{me} inversion 
(vertical axis). For better visualization of low (blue/purple) and high (yellow) density regions, we employ a logarithmic 
color scale. The more the yellow regions are concentrated along the diagonal black line, the better the inference of 
the electric currents is. These plots are in qualitative agreement with those presented in \cite{adur2021currents} in that 
the larger the value of the electric current density, the better the inference, in particular for values $\gtrapprox 0.1$ A m$^{-2}$. 
This feature is seen not only in \emph{me}, but in all tested Stokes inversions.\\

In addition, it is worth noting that, on the low end of the electric current density, the inversion yields an almost constant 
value of $\|\ve{j}_{\rm inv}\| \approx 10^{-1.3} \approx 0.05$ A m$^{-2}$ (see the horizontal thick dashed line in the rightmost panel of 
Fig.~\ref{fig:currents_me_vs_mhd}), even if the electric current in the original MHD simulation is much lower than 
this value. In \cite{adur2021currents} we referred to this value as the ``floor'' value, and it corresponds to the 
smallest value of electric current density that can be inferred by the Stokes inversion. Although this floor is visually 
clearest in the modulus of the electric current density vector, $\|\ve{j}\|$ (rightmost panel in Fig.~\ref{fig:currents_me_vs_mhd}), 
it can been seen in all three individual components of the electric current vector as well. The floor values for all Stokes 
inversions carried out in this work are presented in Table~\ref{tab:floor}. In this table we provide $f$, which we define as 
the power of ten that yields the floor value as $10^{-f}$. With this definition, the larger the value of $f$, the better a 
given inversion type is at inferring smaller values of the electric current density. The numbers in this table were obtained by calculating 
the first moment of the distribution of $\log\|j_{\rm inv}\|$ in the regions where $\|j_{\rm mhd}\| < 0.01$ A~m$^{-2}$ as well as for 
the three individual components of the electric current density vector. According to Table~\ref{tab:floor}, $f$ is smaller for 
$\|j_{\rm z}\|$ (leftmost panel) than for $\|j_{\rm x}\|$ and $\|j_{\rm y}\|$, and therefore the latter components are less affected
by the floor value than the former. This also makes the vertical component of the electric current density, $j_{\rm z}$, the main 
contributor to the floor value of the modulus $\|\ve{j}\|$. The floor values are also slightly smaller at $\tau_c=1$ than at 
$\tau_c=0.1$, meaning that it is possible to detect slightly lower electric currents at the latter optical depth ($\tau_c=0.1$)
than at the former ($\tau_c=1$). Floor values are, however, of limited relevance for our work because such small electric 
current density values play a negligible role in the energy balance of the solar photosphere, as Ohmic heating varies with $\|{\bf j}\|^2$.\\

\begin{figure*}
\begin{tabular}{cc}
   \includegraphics[width=8cm]{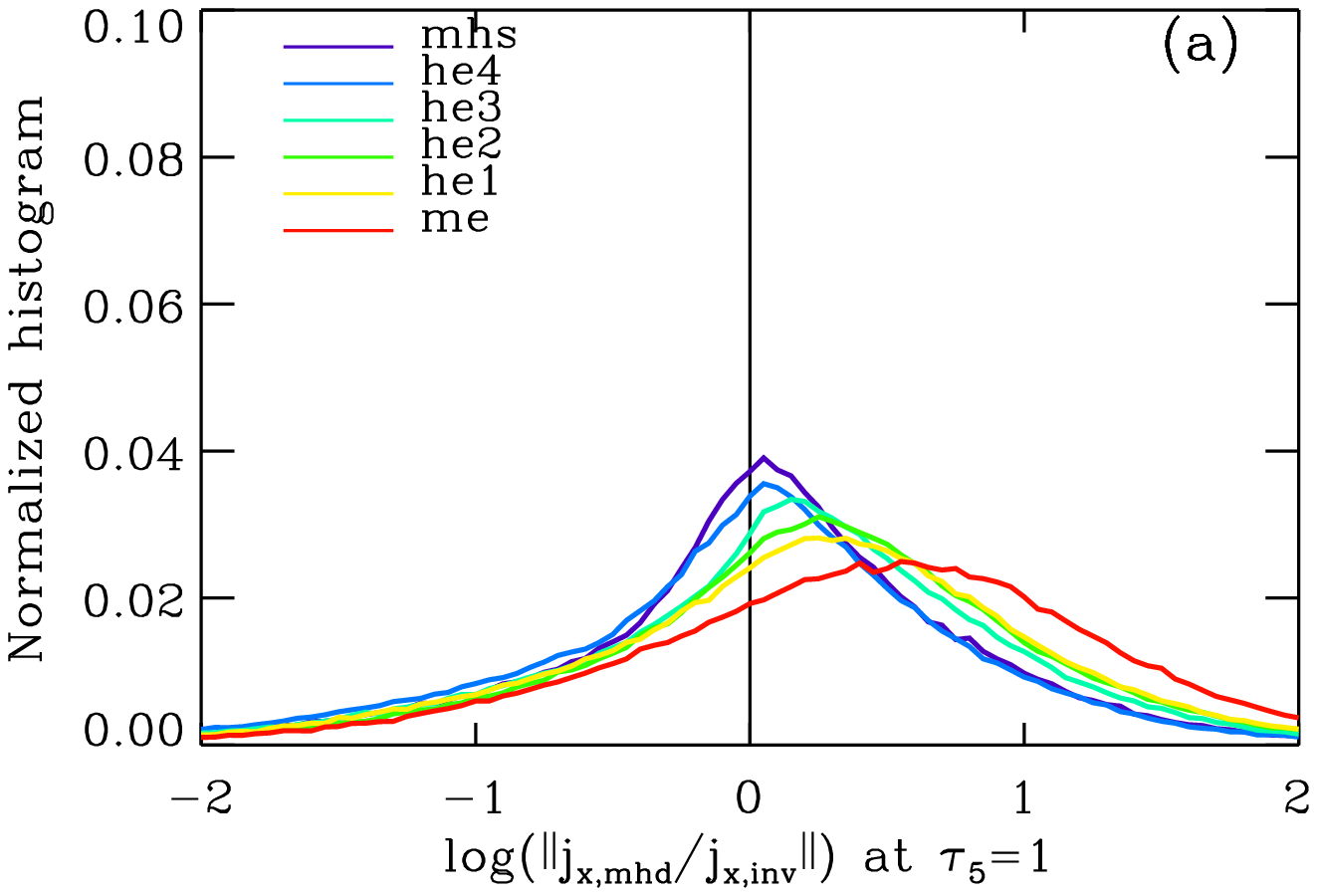} &
   \includegraphics[width=8cm]{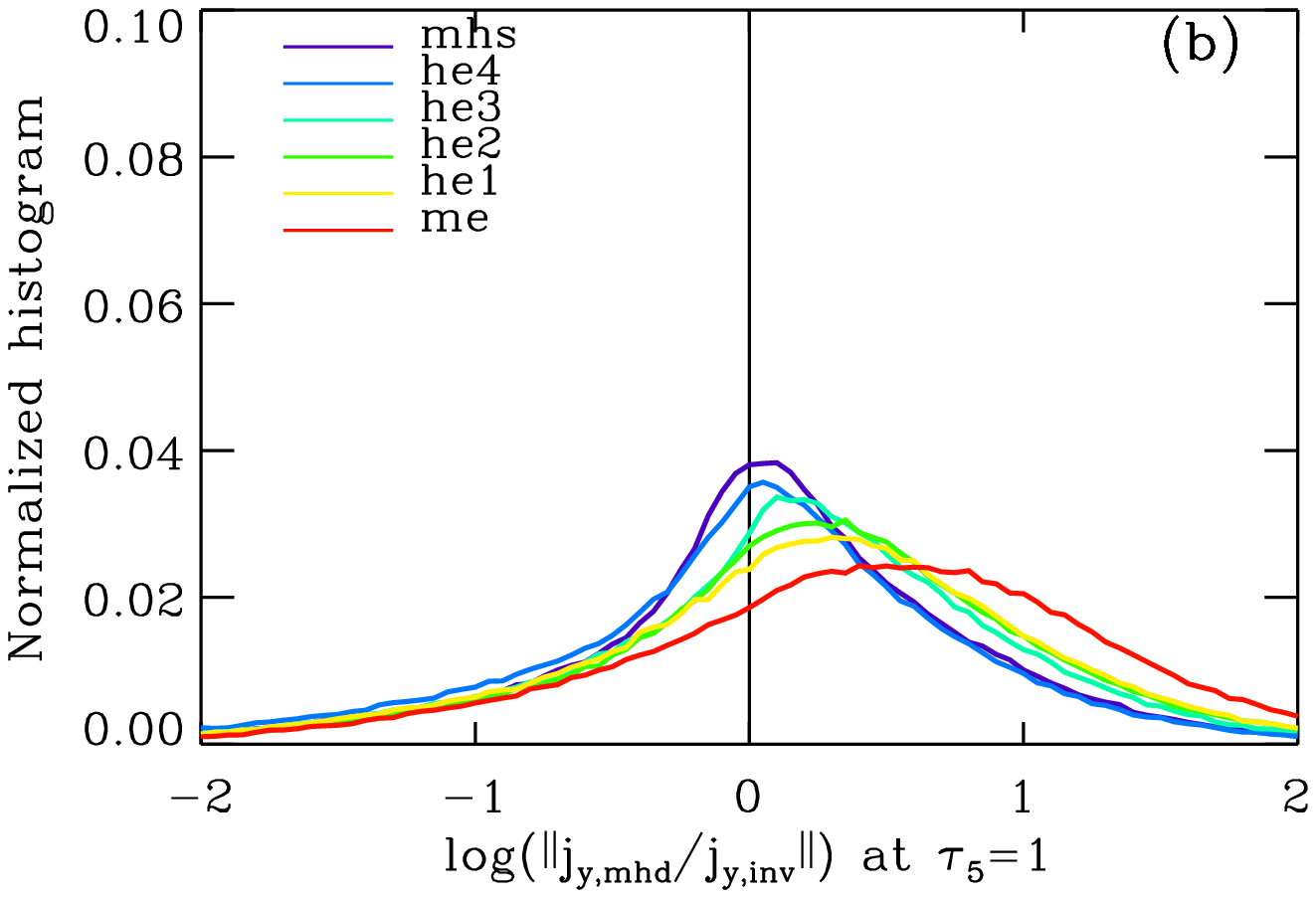} \\
   \includegraphics[width=8cm]{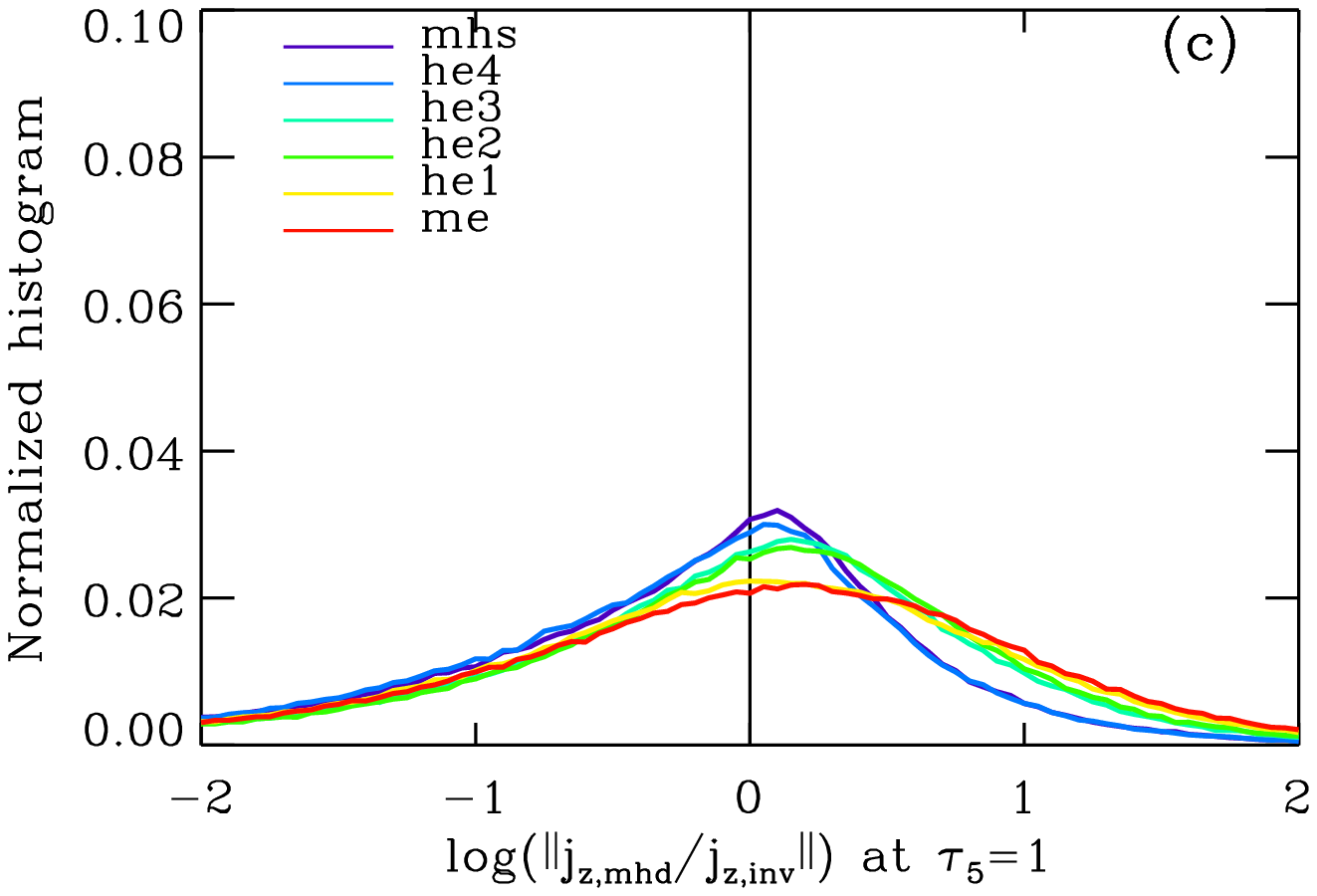} &
   \includegraphics[width=8cm]{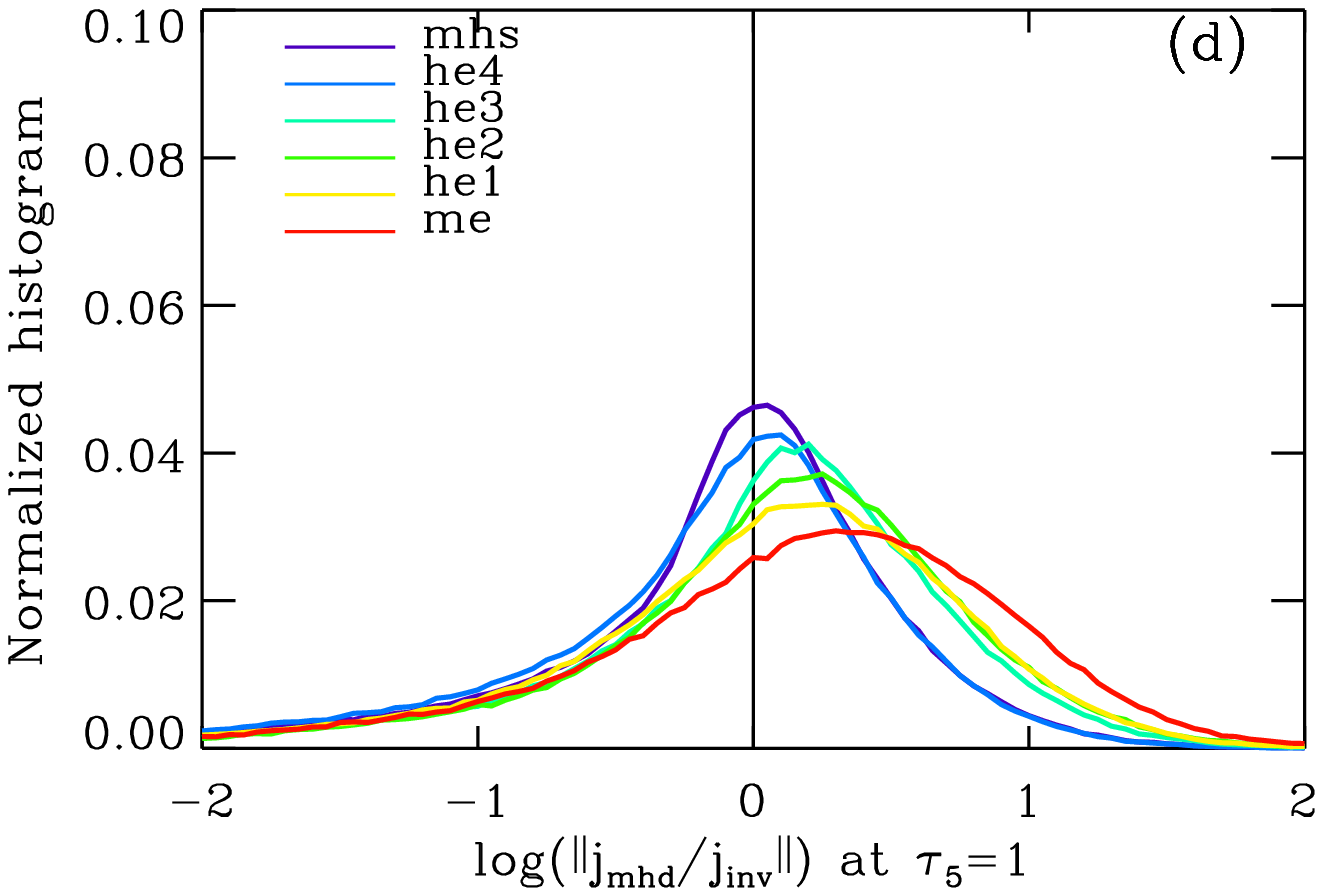} \\
\end{tabular}
   \caption{Histograms of the logarithm of the absolute value of the quotient of the components of the electric current
density from the MHD simulations and as obtained from the Stokes inversion: {\bf (a}) $\|j_{\rm x}\|$ (upper left); 
{\bf (b}) $\|j_{\rm y}\|$ (upper right);  {\bf (c}) $\|j_{\rm z}\|$ (lower left); and {\bf  (d}) $\|\ve{j}\|$ (lower right). The histograms
correspond to the Stokes inversions described in ~Sect.~\ref{sec:inference}: red for \emph{me}, yellow for \emph{he1}, green for \emph{he2},
cyan for \emph{he3}, blue for \emph{he4}, and purple for \emph{mhs}.\label{fig:current_hist1d_all}} 
\end{figure*}

\begin{table}
\caption{Powers of ten, $f$, that yield the ``{floor''} values as $10^{-f}$, at two different optical depths 
($\tau_c=1$,$\tau_c=0.1$) for the different inversions carried out in Sect.~\ref{sec:inference}.} 
\label{tab:floor} 
\centering       
\begin{tabular}{ccccccc}
\hline
 & {\bf me} & {\bf he1} & {\bf he2} & {\bf he3} & {\bf he4} & {\bf mhs} \\
\hline
$\|j_{\rm x}(\tau_c=1)\|$ & 2.10 & 1.85 & 1.95 & 1.73 & 1.56 & 1.76 \\
$\|j_{\rm y}(\tau_c=1)\|$ & 2.08 & 1.85 & 1.96 & 1.80 & 1.57 & 1.77 \\
$\|j_{\rm z}(\tau_c=1)\|$ & 1.37 & 1.36 & 1.49 & 1.48 & 1.11 & 1.14 \\
$\|\ve{j}(\tau_c=1)\|$ & 1.33 & 1.27 & 1.41 & 1.33 & 0.99 & 1.08 \\
\hline
$\|j_{\rm x}(\tau_c=0.1)\|$ & 2.17 & 2.08 & 2.17 & 1.82 & 1.54 & 1.63 \\
$\|j_{\rm y}(\tau_c=0.1)\|$ & 2.13 & 2.07 & 2.16 & 1.84 & 1.57 & 1.69 \\
$\|j_{\rm z}(\tau_c=0.1)\|$ & 1.40 & 1.41 & 1.54 & 1.49 & 1.23 & 1.28 \\
$\|\ve{j}(\tau_c=0.1)\|$ & 1.37 & 1.36 & 1.51 & 1.38 & 1.05 & 1.09 \\
\hline                    
\end{tabular}
\end{table}
A more quantitative analysis can be made via Fig.~\ref{fig:current_hist1d_all}, where we plot histograms of the
logarithm of the absolute value of the ratio between the currents present in the MHD simulations and the currents inferred 
from the Stokes inversion: $j_{\rm x}$ (top left; panel {\bf a}), $j_{\rm y}$ (top right; panel {\bf b}), $j_{\rm z}$
(bottom left; panel {\bf c}), and $\|\ve{j}\|$ (bottom right; panel {\bf d}). Figure~\ref{fig:current_hist1d_all} corresponds to $\tau_c=1$. 
Very similar results are obtained for $\tau_c=0.1$ (not shown). Results for $\tau_c < 10^{-2}$ deteriorate significantly 
(see Sect.~\ref{subsec:height}).\\

Pixels on the positive side of the horizontal axis in Fig.~\ref{fig:current_hist1d_all} correspond to 
regions where the inversion underestimates the true values from the MHD simulations, whereas on the negative side of the 
horizontal axis the inversion overestimates the values from the MHD simulations. All histograms in this figure, in particular 
those corresponding to $j_{\rm z}$ and $\|\ve{j}\|$ (panels {\bf c} and {\bf d}), feature a negative skewness, with a long tail 
on the negative side of the horizontal axis. This tail is produced by the floor value described above, where the inversion 
overestimates the values of the electric current density. In addition, it can be noted that the peak of the different histograms 
is usually shifted toward positive values, meaning that the inversion tends to underestimate the values of the current
density in the MHD simulations. This underestimation happens for values of $\log(\|\ve{j}_{\rm inv}\|) > 0$ A~m$^{-2}$ (see the $\|\ve{j}\|$ retrieval in the rightmost panels in 
Fig.~\ref{fig:currents_me_vs_mhd}). This bias becomes less important
as the complexity of the inversion setup increases, as demonstrated by the fact that the peak of the histograms in
\emph{he4}  and \emph{mhs} are mostly centered at zero (see Fig.~\ref{fig:current_hist1d_all}).\\
\begin{figure*}
\begin{tabular}{cc}
   \includegraphics[width=8cm]{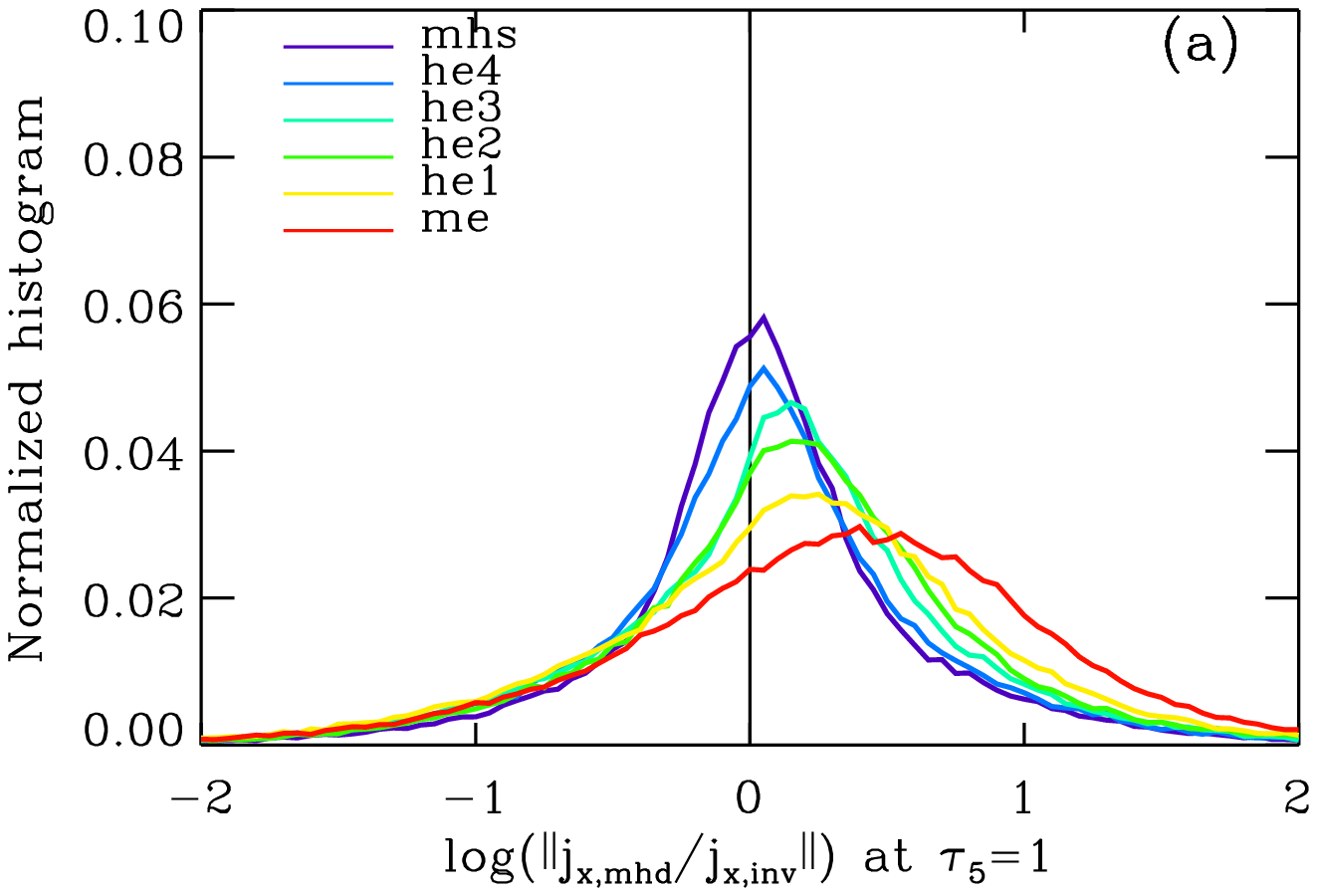} &
   \includegraphics[width=8cm]{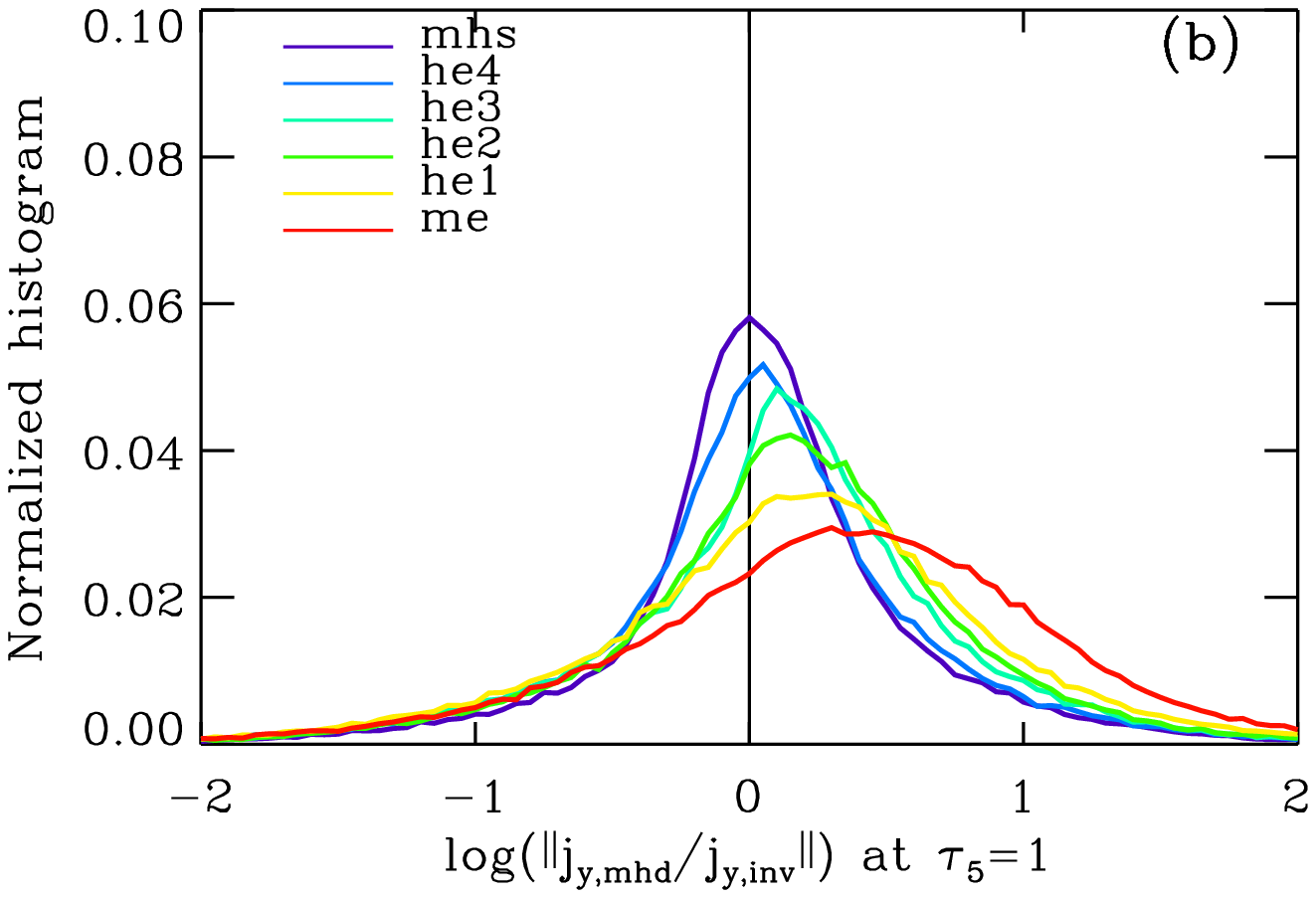} \\
   \includegraphics[width=8cm]{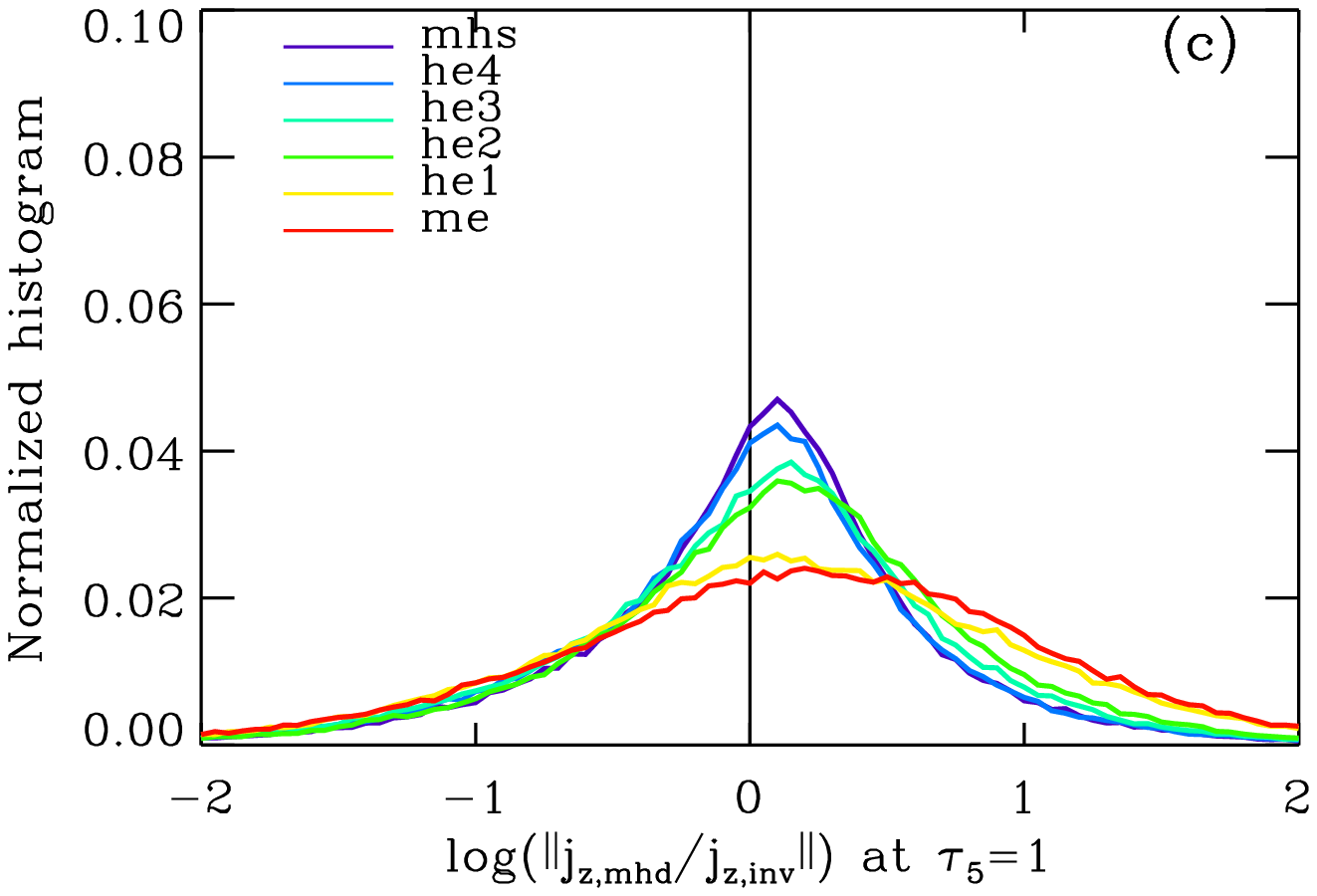} &
   \includegraphics[width=8cm]{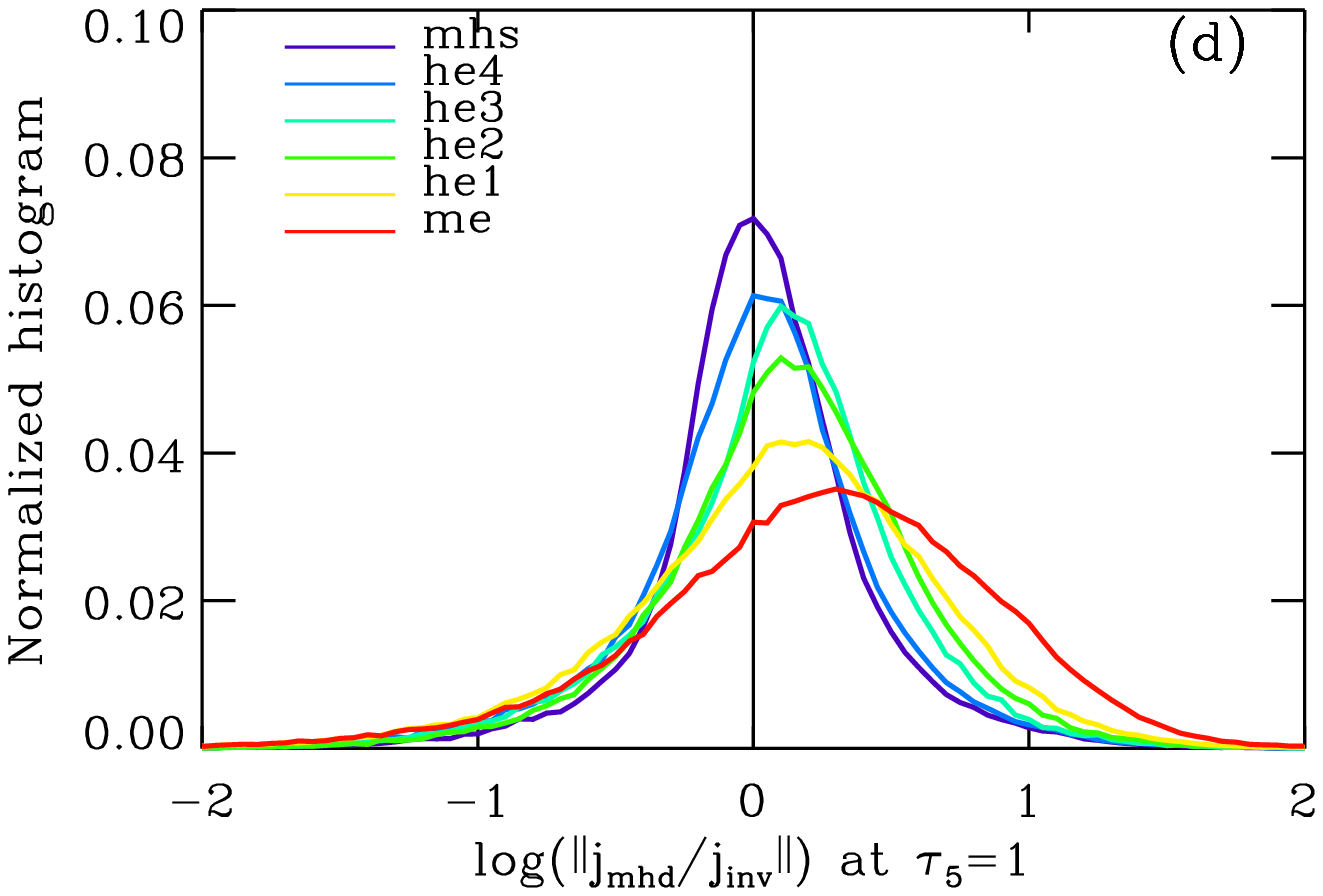} \\
\end{tabular}
   \caption{Same as Fig.~\ref{fig:current_hist1d_all} but employing only regions where $\|\ve{B}(\tau_c=1)\| > 300$ gauss.
\label{fig:current_hist1d_bf}}
\end{figure*}
\begin{table*}
\caption{Values of $\sigma$ calculated from Fig.~\ref{fig:current_hist1d_all} (see text for details)
at two different optical depths ($\tau_c=1,0.1$) for the different Stokes inversions described in Sect.~\ref{sec:inference}.
For each Stokes inversion type, we provide two values (e.g., 0.85/0.70). The first one (e.g., 0.85) corresponds to the standard deviation,
$\sigma$, obtained from the histogram that includes all pixels in the simulations (Fig.~\ref{fig:current_hist1d_all}), while
the second one (e.g., 0.70) is the standard deviation, $\sigma$, obtained from the histogram that includes only pixels where 
$\|\ve{B}(\tau_c=1)\| > 300$ (Fig.~\ref{fig:current_hist1d_bf}).\label{tab:sigma_perce_all}}
\centering       
\begin{tabular}{ccccccc}
\hline
 $\|j_{\rm x}\|$ & {\bf me} & {\bf he1} & {\bf he2} & {\bf he3} & {\bf he4} & {\bf mhs} \\
\hline
$\sigma(\tau_c=1)$ & 1.00/0.84 & 0.85/0.70 & 0.81/0.60 & 0.78/0.58 & 0.73/0.51 & 0.69/0.44 \\
$\sigma(\tau_c=0.1)$ & 0.91/0.84 & 0.90/0.79 & 0.89/0.67 & 0.83/0.66 & 0.86/0.73 & 0.75/0.53 \\
\hline
\hline
 $\|j_{\rm y}\|$ & {\bf me} & {\bf he1} & {\bf he2} & {\bf he3} & {\bf he4} & {\bf mhs} \\
\hline
$\sigma(\tau_c=1)$ & 1.02/0.84 & 0.85/0.69 & 0.82/0.59 & 0.78/0.57 & 0.73/0.51 & 0.69/0.42 \\
$\sigma(\tau_c=0.1)$ & 0.92/0.85 & 0.90/0.77 & 0.90/0.66 & 0.84/0.65 & 0.86/0.72 & 0.73/0.52 \\
\hline
\hline
 $\|j_{\rm z}\|$ & {\bf me} & {\bf he1} & {\bf he2} & {\bf he3} & {\bf he4} & {\bf mhs} \\
\hline
$\sigma(\tau_c=1)$ & 0.98/0.90 & 0.95/0.86 & 0.84/0.65 & 0.82/0.63 & 0.82/0.58 & 0.80/0.54 \\
$\sigma(\tau_c=0.1)$ & 1.03/0.81 & 1.03/0.82 & 0.89/0.60 & 0.89/0.62 & 0.95/0.63 & 0.90/0.57 \\
\hline
\hline
 $\|\ve{j}\|$ & {\bf me} & {\bf he1} & {\bf he2} & {\bf he3} & {\bf he4} & {\bf mhs} \\
\hline
$\sigma(\tau_c=1)$ & 0.77/0.64 & 0.65/0.51 & 0.61/0.42 & 0.57/0.39 & 0.51/0.33 & 0.48/0.28 \\
$\sigma(\tau_c=0.1)$ & 0.74/0.64 & 0.73/0.57 & 0.70/0.44 & 0.65/0.45 & 0.66/0.49 & 0.56/0.34 \\
\hline
\hline
\end{tabular}
\end{table*}

In order to evaluate the ability of the different inversions to retrieve the correct values of electric currents in the
MHD simulations, we considered the standard deviation, $\sigma$, of the histograms \footnote{This $\sigma$ here refers to the standard
deviation of the retrieval of the electric current density. It is not to be confused with the $\sigma$ in Table~\ref{tab:spectral_ranges},
which refers to the cross section for collisions for the spectral lines used in this work. It is also different from $\sigma_{pn}$, which
refers to the standard deviation of the photon noise in units of the continuum intensity (see Sect.~\ref{sec:synthesis}).}. The smaller the value of $\sigma$, the better 
the retrieval of the corresponding component of the electric current density. Values of $\sigma$ at two different optical depths, 
$\tau_c=0.1,1$, in all considered inversions are presented in Table~\ref{tab:sigma_perce_all}. For each inversion we provide two values of $\sigma$
as $XXX/YYY$. The first value, $XXX$, corresponds to the standard deviation of the histogram in all pixels of the simulation (Fig.~\ref{fig:current_hist1d_all}), 
whereas the second value, $YYY$, is the standard deviation obtained from the histograms using only those pixels where the magnetic field is stronger than 300 gauss
(Fig.~\ref{fig:current_hist1d_bf}). This second estimation of $\sigma$ is provided because, as pointed out in \citet{adur2021currents}, the electric current 
density is determined much more reliably in regions that harbor stronger magnetic fields. We emphasize here that the values given for the standard deviation of the histograms 
in Table~\ref{tab:sigma_perce_all} are given for $\log[\|j_{\rm mhd}/j_{\rm inv}\|]$. This means that a standard deviation of, for example, 0.7 dex implies that the values of the electric 
current inferred through the inversion are, about 68~\% of the time, between $j_{\rm inv} \in [10^{-0.7}, 10^{0.7}] \approx [1/5,5] j_{\rm mhd}$. In other words, it implies that the 
electric current can be inferred within a factor of 5 of the original values (higher or lower)  in the MHD simulations.\\

An analysis of Table~\ref{tab:sigma_perce_all} reveals that the values of $\sigma$ typically decrease from \emph{me} to \emph{mhs}. This implies that the inference of all three 
components of the electric current density improves as the complexity of the performed Stokes inversion increases. The overall 
improvement in $\sigma$ from the simplest to the most complex inversion is about 25-50 \%. This is not due to a particular inversion
setup; they each produce a similar and steady enhancement in the accuracy of the retrieval of the components of the electric current 
density. The only notable exception seems to be \emph{he4}, where, as anticipated in Sect.~\ref{subsec:regul}, inferences of $j_z$ at $\tau_c=0.1$
worsen with respect to \emph{he3}. Table~\ref{tab:sigma_perce_all} also confirms that the retrieval of $j_{\rm z}$ is always 
less accurate than for $j_{\rm x}$ and $j_{\rm y}$ and that currents can be determined more reliably at $\tau_c=1$ than at $\tau_c=0.1$, although at 
$\tau_c=0.1$ it is possible to detect lower values of the components of the electric current density (see Table~\ref{tab:floor}).\\

If we include all pixels in the simulation (using only the first number in Fig.~\ref{fig:current_hist1d_all}), we can summarize our results
by saying that, under the simplest Stokes inversion described in Sect.~\ref{sec:inference} (\emph{me}), the three components of the electric current density can 
be determined with a standard deviation of $\sigma=0.90-1.00$ dex. In the case of the most advanced Stokes inversion (\emph{mhs}), the three components of 
the electric currents can be determined with an accuracy of $\sigma=0.70-0.85$ dex. The total electric current density can be retrieved with $\sigma=0.75$ dex
for \emph{me} and $\sigma=0.50$ dex for \emph{mhs}.\\

Table~\ref{tab:sigma_perce_all} also shows that the inference of the electric current density clearly improves when only regions of strong magnetic fields are 
considered ($\|\ve{B}(\tau_c=1)\| > 300$ gauss). This can be readily seen by comparing the widths and heights of the histograms 
in Fig.~\ref{fig:current_hist1d_bf} with those in Fig.~\ref{fig:current_hist1d_all}. Results in this case can be summarized as follows. Under \emph{me}, 
the three components of the electric current density can be determined with a standard deviation of $\sigma=0.85-0.90$ dex, whereas for \emph{mhs} $\sigma=0.5-0.6$ dex.
When referring to the total current density, the accuracy ranges between $\sigma=0.65$ dex for \emph{he1} and $\sigma=0.30$  dex for \emph{mhs}.\\

\begin{figure*}
\centering
   \includegraphics[width=18cm]{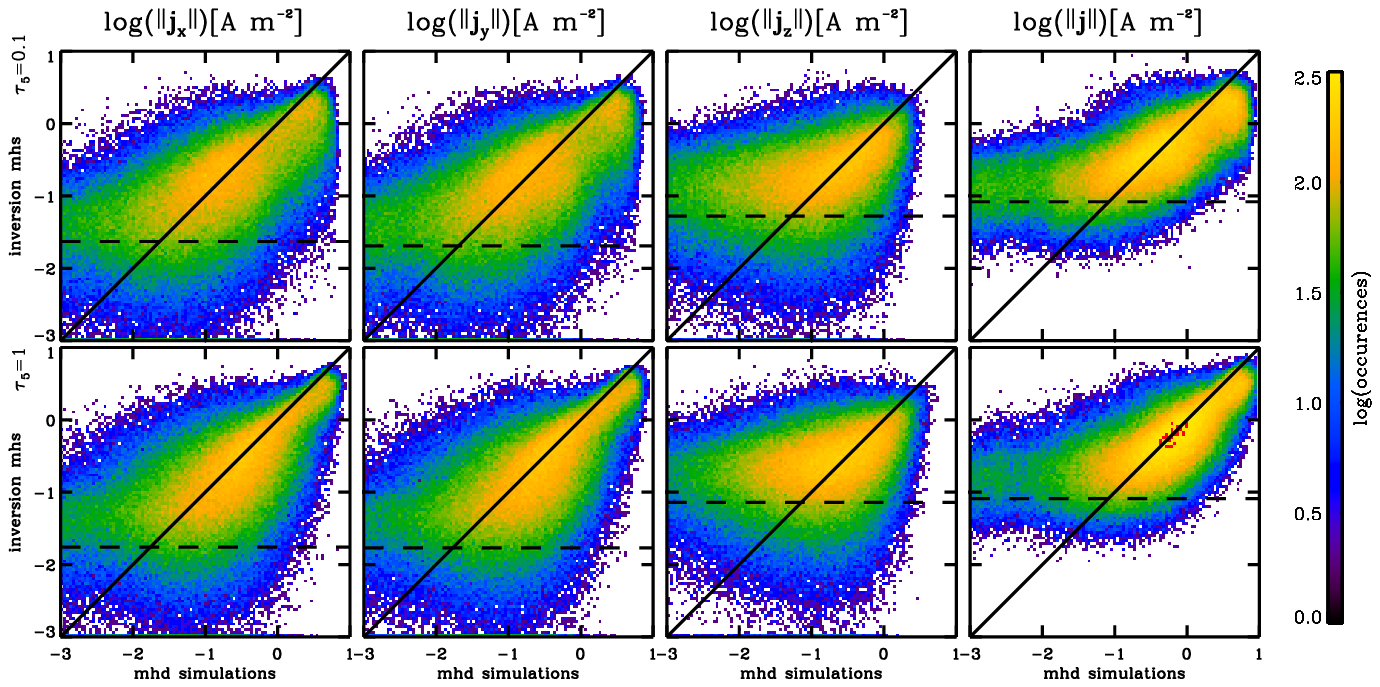}
   \caption{Same as Fig.~\ref{fig:currents_me_vs_mhd} but showing the electric currents as obtained from the inversions under 
MHS equilibrium described in Sect.~\ref{subsec:mhs} (\emph{mhs}).\label{fig:currents_mhs_vs_mhd}} 
\end{figure*}

A visual impression of the overall improvement between \emph{me} and \emph{mhs} can be drawn by comparing Fig.~\ref{fig:currents_me_vs_mhd} with 
Fig.~\ref{fig:currents_mhs_vs_mhd}, where we present the logarithmic scatter-density plots of the electric current density from the MHD 
simulations (horizontal axis) and the values inferred by the \emph{mhs} Stokes inversion (vertical axis) at $\tau_c=1$ (lower panels) and 
$\tau_c=0.1$ (upper panels). These two figures illustrate the improvement between the simplest and the most complex inversion that we carried
out. However, we emphasize that the improvement is not solely due to the use of MHS equilibrium (as opposed to HE) but also the
cumulative effect of all the inversions described in Sect.~\ref{sec:inference}. \\

\subsection{Spatial derivatives}
\label{subsec:deriv}

In Sect.~\ref{subsec:summary} we have seen that $j_{\rm x}$ and $j_{\rm y}$ are always better retrieved than $j_{\rm z}$. In order to
understand the reason, we need to investigate our ability to infer the individual spatial derivatives that enter the calculation of the three
components of the electric current density: 

\begin{eqnarray}
j_{\rm x} & \propto (\partial B_{\rm y}/\partial z) - (\partial B_{\rm z}/\partial y)\notag\\
j_{\rm y} & \propto (\partial B_{\rm x}/\partial z) - (\partial B_{\rm z}/\partial x)\\
j_{\rm z} & \propto (\partial B_{\rm x}/\partial y) - (\partial B_{\rm y}/\partial x)\notag
\label{eq:j}.
\end{eqnarray}

\begin{figure}
  \includegraphics[width=8cm]{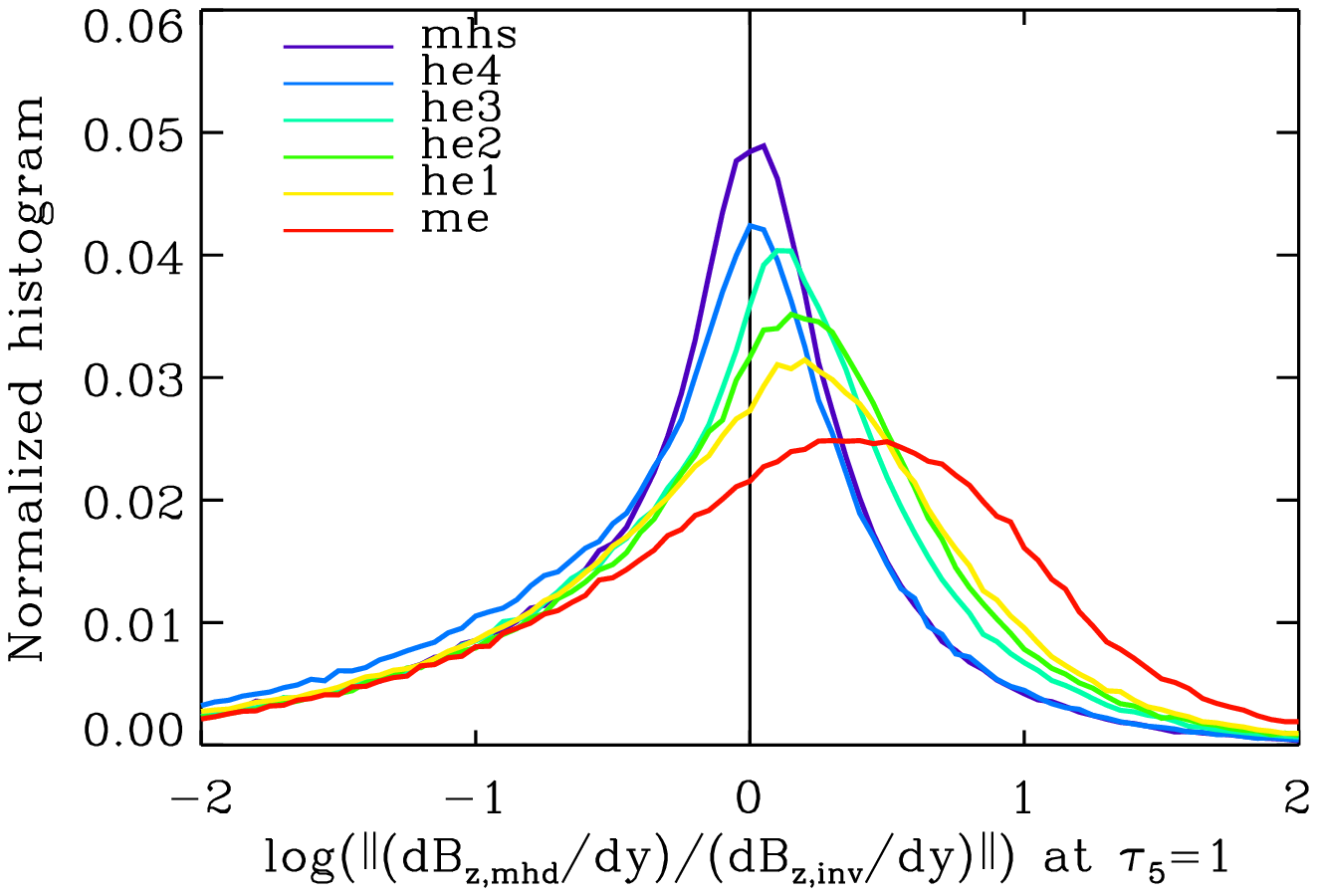} \\
  \includegraphics[width=8cm]{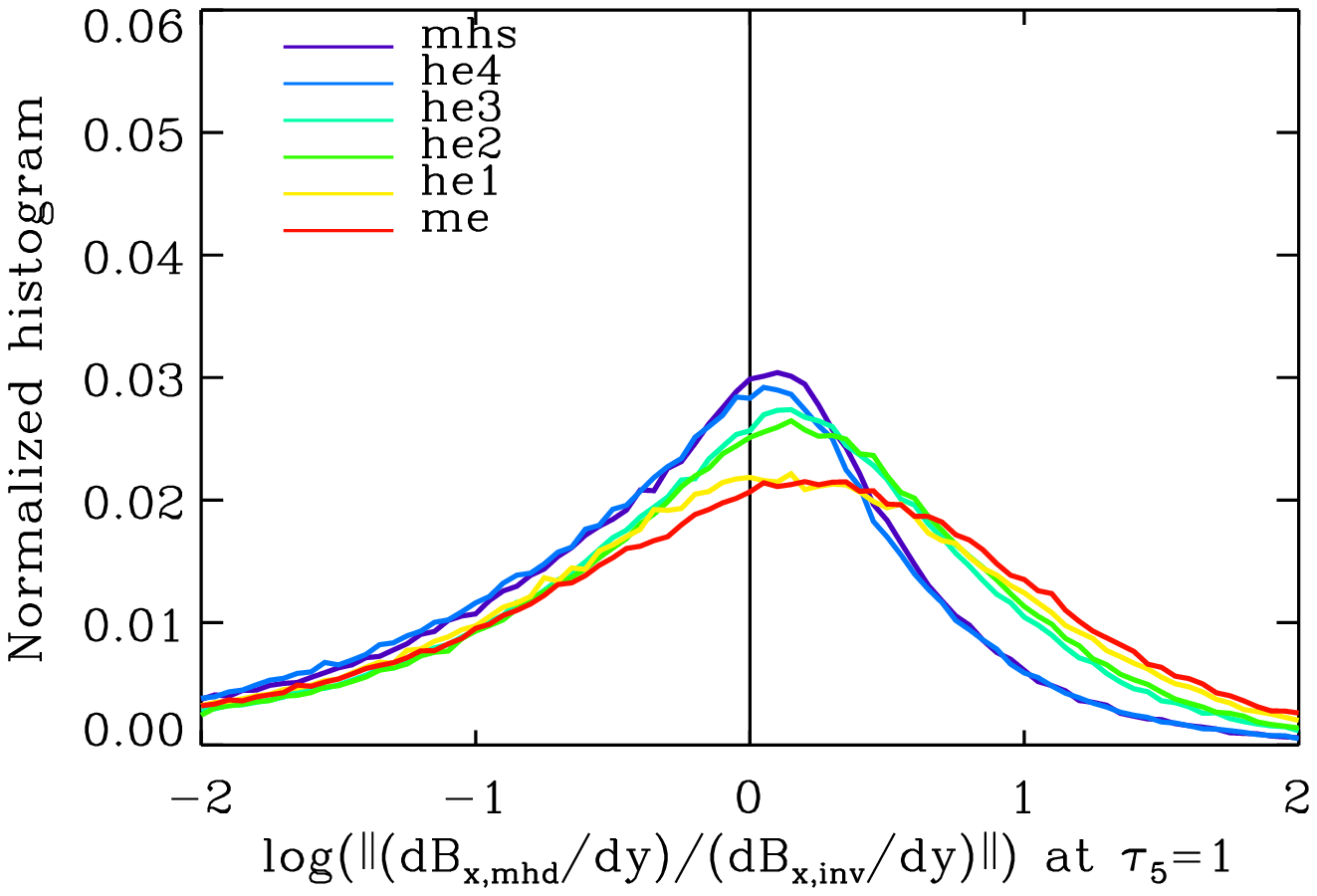} \\
  \includegraphics[width=8cm]{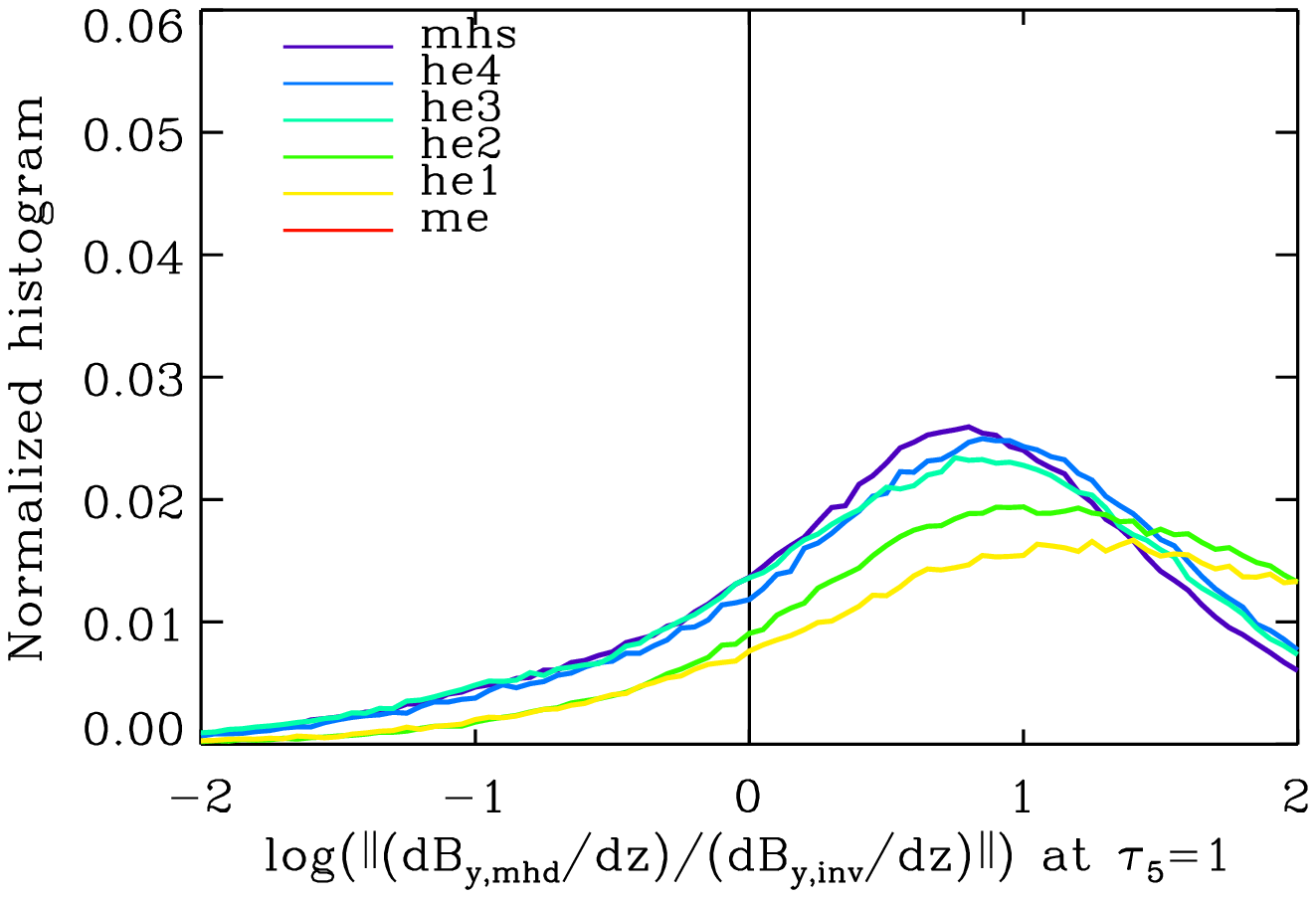}
   \caption{Histograms of the logarithm of the absolute value of the quotient of the spatial derivatives of the components 
of the electric current density from the MHD simulations and as obtained from the Stokes inversion. {\it Top panel}: $\partial B_{\rm z}/\partial y$ 
(or alternatively $\partial B_{\rm z}/\partial x$). {\it Middle panel}: $\partial B_{\rm x}/\partial y$ (or alternatively $\partial B_{\rm y}/\partial x$). 
{\it Bottom panel}: $\partial B_{\rm y}/\partial z$ (or alternatively $\partial B_{\rm x}/\partial z$). The histograms correspond to the Stokes inversions 
described in ~Sect.~\ref{sec:inference}: \emph{me} (red), \emph{he1} (yellow), \emph{he2} (green), \emph{he3} (cyan), \emph{he4} (blue), and \emph{mhs} (purple).\label{fig:deriv}}
\end{figure}

Figure~\ref{fig:deriv} shows histograms of the logarithm of the ratio between the derivatives in the MHD simulations and the derivatives as
inferred from the different Stokes inversions described in Sect.~\ref{sec:inference}. The histograms are similar to those in Fig.~\ref{fig:current_hist1d_all}
and so are their interpretation: the positive region on the horizontal axis corresponds to pixels where
the spatial derivative in the MHD simulation is larger than that inferred through the inversion, whereas the negative region on the horizontal axis corresponds to pixels
where the spatial derivative inferred through the inversion is larger than the derivative in the MHD simulations. In this figure we show only three of the six
derivatives in Eq.~\ref{eq:j} because, statistically speaking, $\partial B_{\rm z}/\partial y$ (top panel) behaves identically to $\partial B_{\rm z}/\partial x$,
$\partial B_{\rm x}/\partial y$ (middle panel) behaves identically to $\partial B_{\rm y}/\partial x$, and finally $\partial B_{\rm y}/\partial z$ (bottom panel) 
behaves identically to $\partial B_{\rm x}/\partial z$. Therefore, these three panels suffice for the argumentation that follows.\\

The first obvious conclusion we can draw from Fig.~\ref{fig:deriv} is that, just as we saw with the electric current density, the inference of the spatial derivatives 
of the magnetic field steadily improves as the complexity of the inversion increases. Histograms are narrower and more centered at the zero value as we go from \emph{me} (Sect.~\ref{subsec:me})
to \emph{mhs} (Sect.~\ref{subsec:mhs}). The second conclusion is that, from best to worst, the order in which spatial derivatives are best inferred is: 
$\partial B_{\rm z}/\partial x$ (or $\partial B_{\rm z}/\partial y$; top panel in Fig.~\ref{fig:deriv}), $\partial B_{\rm x}/\partial y$ (or $\partial B_{\rm y}/\partial x$; middle panel), and finally $\partial B_{\rm y}/\partial z$ (or $\partial B_{\rm x}/\partial z$; bottom panel). The reason as to why this is the case can be
understood in terms of the sensitivity of the three components of the magnetic field to the different Stokes profiles. The $B_{\rm z}$ is generally better retrieved
than either $B_{\rm x}$ or $B_{\rm y}$ because signals in Stokes $V$ are typically larger than in $Q$ and $U$ even if the vertical and horizontal components of the
magnetic field have the same value \citep[see, e.g., Fig.~5 in][]{borrero2013qs}. This explains why $\partial B_{\rm z}/\partial x$ is better determined than both $\partial B_{\rm x}/\partial y$ and 
$\partial B_{\rm x}/\partial z$. Now, $\partial B_{\rm x}/\partial y$ is more reliably inferred than $\partial B_{\rm x}/\partial z$ because for the former
we employ data (i.e., the Stokes vector) from two different adjacent pixels, whereas in the latter we employ the data from just one. The increased amount of
information available allows the horizontal derivative of the horizontal components of the magnetic field  to be determined more accurately than the vertical derivative
of the horizontal components of the magnetic field.\\

We switch our attention now to Eq.~\ref{eq:j}. Here we see that $j_{\rm x}$ and $j_{\rm y}$ are obtained from a combination of two spatial derivatives, one of which can be determined very accurately
($\partial B_{\rm z}/\partial y$ or $\partial B_{\rm z}/\partial x$) while the other is rather poorly determined ($\partial B_{\rm y}/\partial z$ or $\partial B_{\rm x}/\partial z$) and is
actually not retrieved at all in the ME-like case (Sect.~\ref{subsec:me}). Meanwhile, $j_{\rm z}$ is obtained from the combination of two spatial derivatives ($\partial B_{\rm x}/\partial y$ and 
$\partial B_{\rm y}/\partial x$), both of which can be determined with an accuracy that is somewhere in between the derivatives needed to determine $j_x$ and $j_y$. This raises the question as 
to what makes the inference of $j_{\rm x}$ and $j_{\rm y}$ more reliable than $j_{\rm z}$. We believe there are two main reasons. To begin with, we find that in the original MHD 
simulations (Sect.~\ref{sec:synthesis}), at $\tau_c=1$ and $\tau_c=0.1$, $\|\partial B_{\rm z}/\partial y\| > \|\partial B_{\rm y}/\partial z\|$ in 55 \% of the analyzed horizontal area. Therefore, for the correct retrieval of $j_{\rm x}$, it is slightly more important to accurately 
determine the first of these two spatial derivatives, which we actually do. An identical argument can be made for the case of $j_{\rm y}$ and its two spatial derivatives 
(Eq.~\ref{eq:j}). Interestingly, this discrepancy does not appear between $\|\partial B_{\rm x}/\partial y\|$ and $\|\partial B_{\rm y}/\partial x\|$: neither one dominates over the other,
and therefore $j_{\rm z}$ is not affected.\\

In addition, it can be seen that the inversion tends to overestimate $\|\partial B_{\rm z}/\partial y\|$, as indicated by the long tail toward negative values in the top panel of Fig.~\ref{fig:deriv}. 
At the same time, the inversion  tends to underestimate $\|\partial B_{\rm y}/\partial z\|$, as demonstrated by the fact that the peak of the histogram in the bottom panel of Fig.~\ref{fig:deriv} always
lies on the positive side. Because of this, the overestimation of one derivative involved in the calculation of $j_{\rm x}$ (and of $j_{\rm y}$) partially compensates for the underestimation
of the other derivative, such that a final, accurate value is yielded.\\
\subsection{Height dependence}
\label{subsec:height}

\begin{figure}
  \includegraphics[width=8cm]{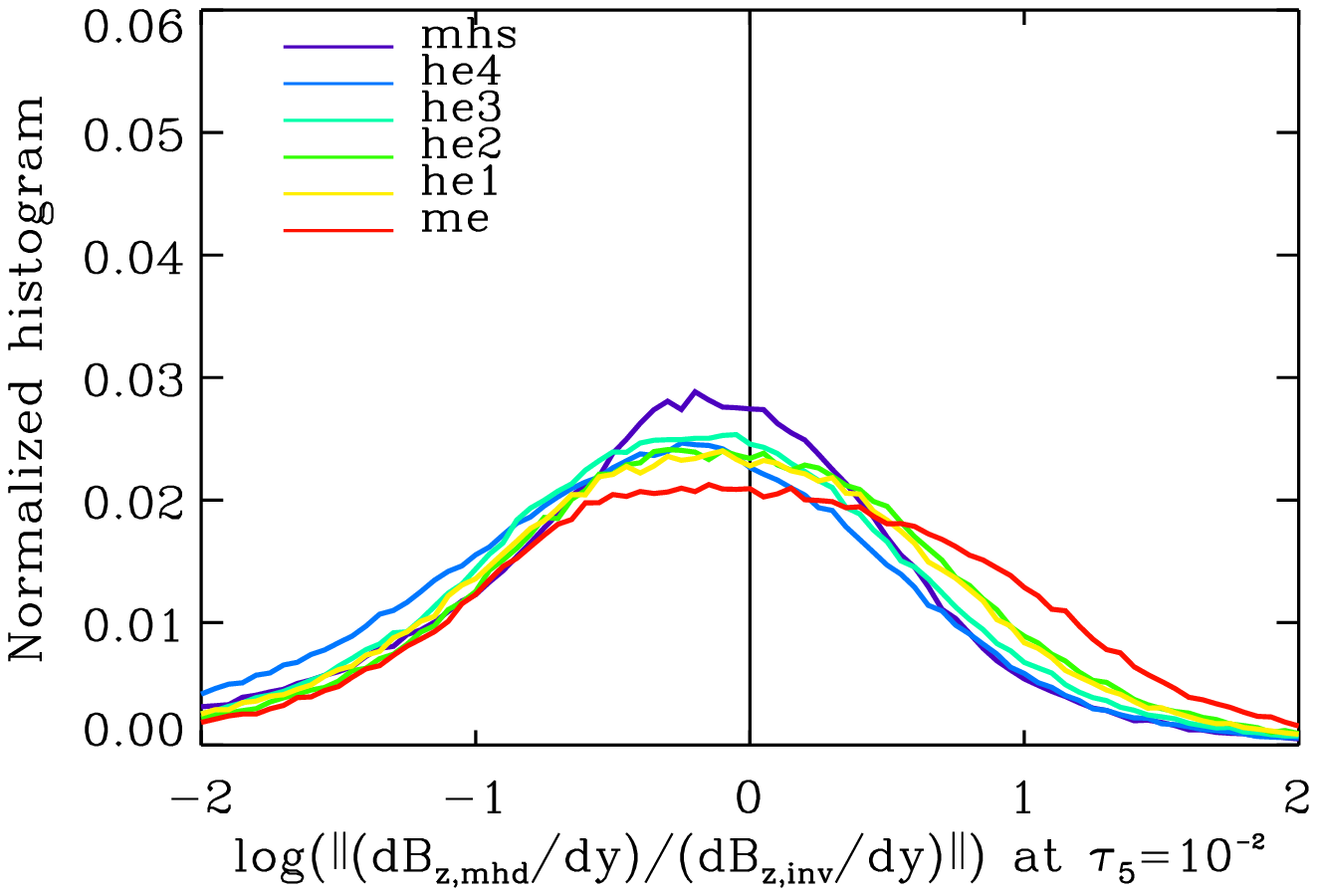} \\
  \includegraphics[width=8cm]{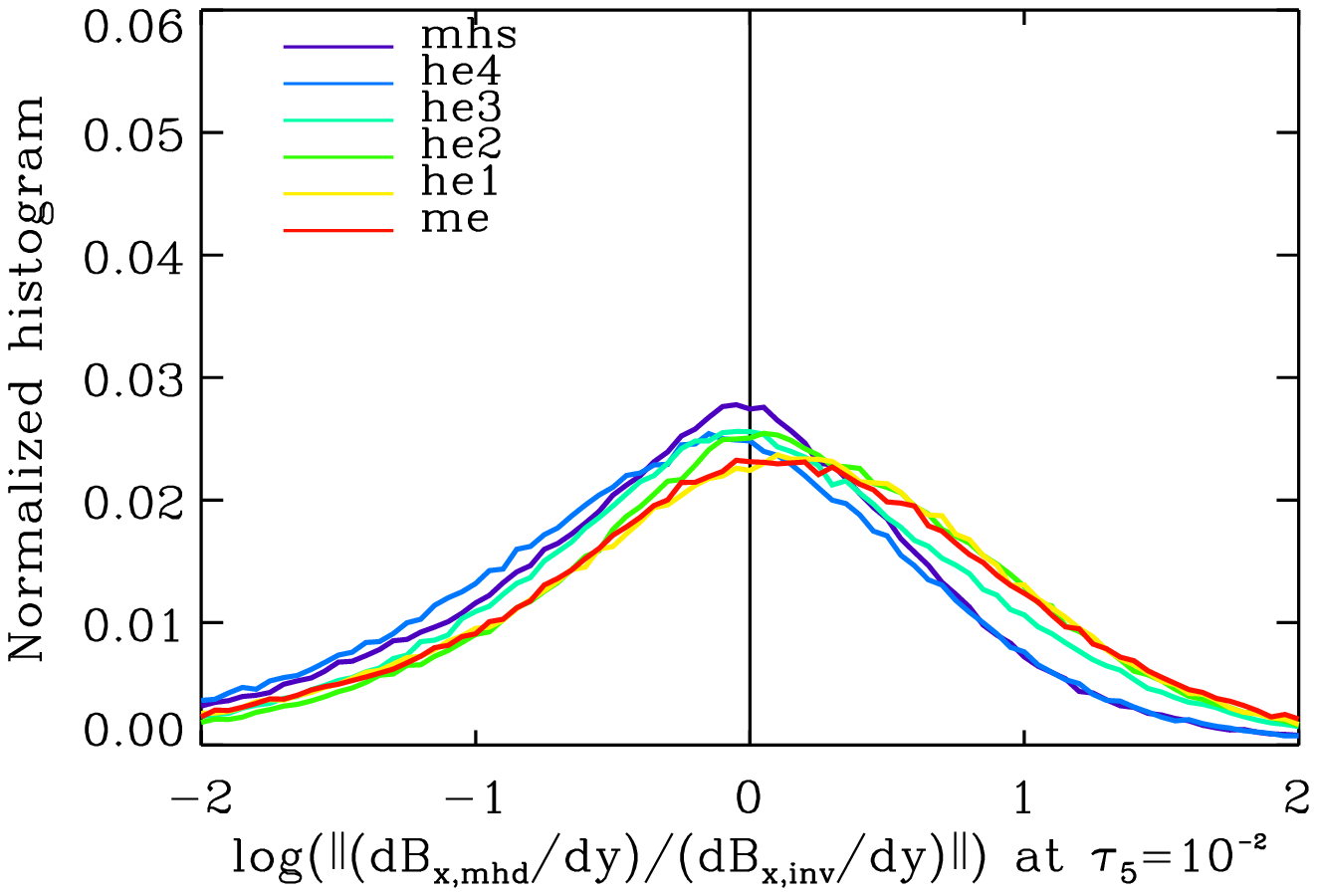} \\
  \includegraphics[width=8cm]{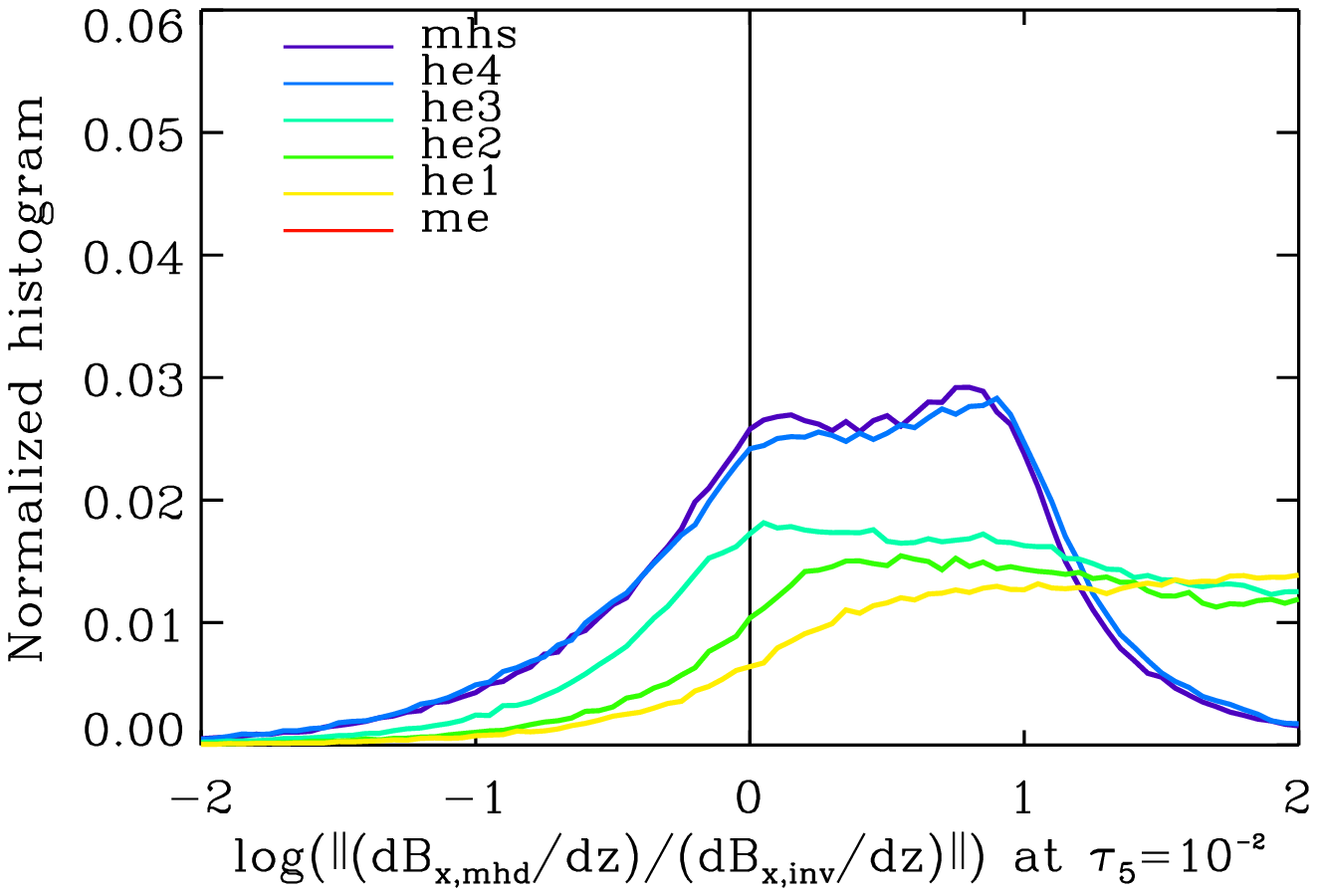}
   \caption{Same as Fig.~\ref{fig:deriv} but at $\tau_c=10^{-2}$.\label{fig:deriv-2}}
\end{figure}

In Sect.~\ref{subsec:summary} we pointed out that the inference of electric currents quickly worsens for optical
depths $\tau_c < 10^{-2}$. We can offer two different explanations as to why this happens. The first is that for $\tau_c < 10^{-2}$ (above the mid-photosphere)
the determination of the spatial derivatives is worse than for $\tau_c \in [0.1,1]$. This is illustrated in Fig.~\ref{fig:deriv-2}, which is to be
compared with Fig.~\ref{fig:deriv}. Unreliable results for the mid-photosphere and above might be caused by the spectral ranges
selected for our analysis (Table~\ref{tab:spectral_ranges}). Our original intention was to include lines with sensitivity to the
magnetic field in a wide variety of heights and optical depths \citep[see Fig.10 in][]{borrero2016pen}. In particular, this was the rationale for including 
Si {\sc I} 1082.7 nm. This was guided by the fact that, for the temperature $T(z)$ and line-of-sight velocity $v_z(z)$, this spectral line features a strong sensitivity 
to the mid- and upper-photosphere \citep{bard2008si,sergio2020si}. Unfortunately, this does not seem to be the case for the magnetic field,
as the sensitivity peaks at $\tau_c=10^{-2}$ and quickly drops for smaller optical depths \citep{kuckein2012si,tobias2016si,natalia2019si}.\\

The second reason is that, even if the spatial derivatives of the magnetic field could be determined for $\tau_c=10^{-2}$ (Fig.~\ref{fig:deriv-2}) 
as accurately as for $\tau_c=1$ (Fig.~\ref{fig:deriv}), the determination of the components of the electric current density vector would still worsen because, 
unlike at $\tau_c=1$ (where $\|\partial B_{\rm z}/\partial y\| > \|\partial B_{\rm y}/\partial z\|$ in 55 \% of the analyzed area), at $\tau_c=10^{-2}$ the opposite happens: 
$\|\partial B_{\rm z}/\partial y\| < \|\partial B_{\rm y}/\partial z\|$ in about 65 \% of the area. And consequently, since $\|\partial B_{\rm y}/\partial z\|$ is less accurate 
than $\|\partial B_{\rm z}/\partial y\|$, the resulting $j_{\rm x}$ would also be less reliable.\\

The tendency for $\|\partial B_{\rm z}/\partial y\| < \|\partial B_{\rm y}/\partial z\|$ also continues at $\tau_c=10^{-3}$, but higher up,
at $\tau_c=10^{-4}$, neither dominates over the other. Unfortunately, at this height the sensitivity to the magnetic field
provided by the selected spectral ranges in Table~\ref{tab:spectral_ranges} is already negligible. Moreover, at this height it is not even clear
if the MHS approach employed here to determine the gas pressure and the $z$ scale (Sect.~\ref{subsec:mhs}) is valid. The reason is that
at $\tau_c < 10^{-4}$, plasma velocities are comparable to the sound speed, and therefore the role of the advection term in the momentum equation 
cannot be neglected. In this case it would be more appropriate to employ a magneto-hydro-stationary approach instead, which unfortunately has not yet been
developed.\\

It is worth mentioning that the fact that some derivatives dominate over others at different heights is a property of the MHD simulations
and of the selected region within the entire simulation domain. We believe this region is large enough to offer a good statistical sample of what
actually happens in the solar atmosphere. However, the real Sun (or different regions on the Sun) might behave somewhat differently. In this
regard, our study does not offer a full answer. Instead, what our analysis does is to establish under which circumstances the electric current density can be correctly
determined in the solar photosphere.\\

\section{Effects of photon noise and spectral resolution}
\label{sec:noise_spec}
In order to study the effect of photon noise, we set up a new inversion in which we analyzed only the spectral lines 
in the first spectral range in Table~\ref{tab:spectral_ranges} without vertical regularization,
using HE, and employing four, four, and eight nodes for $B_{\rm x}(z)$, $B_{\rm y}(z)$, and $B_{\rm z}(z)$, respectively. 
Three different levels of photon noise were employed: $\sigma_p=0$, $\sigma_p=3 \times 10^{-4}$,
and $\sigma_p=10^{-3}$ (values given in units of the quiet Sun continuum intensity). We refer to these inversions as \emph{he5}, \emph{he6}, and \emph{he1}, 
respectively. We note that the third of these inversions, with $\sigma_p=10^{-3}$, was already described in Sect.~\ref{subsec:hydeq}. Results
indicate that as the level of noise increases from \emph{he5}, to \emph{he6,} and to \emph{he1}, the floor values, $f$, become smaller (see Table~\ref{tab:floor_noise_spec}). 
This means that as the noise increases, the smallest value of the electric current that the inversion can detect increases: \emph{he5} can detect 
electric current densities, $\|{\bf j}\|$, as small as $10^{-1.53} \approx 0.029$ A m$^{-2}$, whereas the smallest electric current density that \emph{he1} can detect is 
$10^{-1.36} \approx 0.043$. The largest increase is seen in $j_{\rm z}$, thus indicating that the noise mostly affects the linear polarization profiles
$Q$ and $U$ and thereby the determination of $B_{\rm x}$ and $B_{\rm y}$ \citep{borrero2011qs,borrero2012qs}. Interestingly, the values of $\sigma$ do not 
significantly change as the noise increases (see Table~\ref{tab:sigma_perce_noise_spec}).\\

The fact that the noise does not worsen the determination of the electric current density and the fact that the noiseless case (\emph{he5}) still yields no clear 
improvement points at the model's limitations are the main sources of errors in the determination of the electric currents. By limitations
of the model, we refer to the number of nodes, the interpolation method employed between those nodes, and whether they are sufficient to reproduce the $z$ variations 
of the physical parameters that are present in the MHD simulations. In addition, we need to consider the limitations of radiative transfer itself, whereby variations along 
the line of sight (i.e., in the $z$ direction) of the physical parameters at scales much smaller than the mean-free path of the photons at a given wavelength do not significantly 
influence the emerging Stokes spectra \citep{jc2016review}. This in turn explains why the vertical derivatives of the components of the magnetic field are retrieved 
the worst (see the bottom panel in Fig.~\ref{fig:deriv}). This interpretation is further reinforced by the fact that the more complexity (nodes or free parameters) we allow 
during the inversion, the better the retrieval of the electric current density is.\\

We also studied the effect of spectral resolution by performing an inversion that is identical to \emph{he1}
but convolving the data across the spectral dimension with a transmission profile characterized by a Gaussian function
with a full width half maximum (FWHM) of 40 m{\AA} and later with a FWHM of 80 m{\AA}. These new inversions are referred to as \emph{he7}
and \emph{he8}. The spectral sampling was still 20 m{\AA}~pixel$^{-1}$ (because only the first spectral range in Table~\ref{tab:spectral_ranges}
was inverted). We note that, although \emph{he1} has an infinite spectral resolution (i.e., the convolving function can be viewed as
a zero-width Dirac $\delta$), \emph{he7} and \emph{he8} possess a spectral resolution of 0.025 and 0.0125 m{\AA}$^{-1}$, respectively 
(i.e., inversely proportional to the width of the convolving function; see the footnote in Sect.~\ref{sec:synthesis}). Table~\ref{tab:floor_noise_spec}
shows that the floor values barely change as the spectral resolution degrades. The only noticeable change occurs in the floor value for the 
$j_{\rm z}$ component of the electric current density, which becomes slightly larger ($f=1.36,1.41,1.42$) as the spectral resolution degrades. In this
way, the \emph{he1} inversion allows values of $j_{\rm z}$ as low as $10^{-1.37} \approx 0.042$ A m$^{-2}$ to be determined, and \emph{he8} allows values as low as $10^{-1.42} \approx 0.038$ A m$^{-2}$ to be
inferred. A similar effect was seen in Sect.~\ref{subsec:summary}, where the least complex
inversion (\emph{he1}) allowed the lowest values of the electric current density to be correctly inferred. Table~\ref{tab:sigma_perce_noise_spec} also shows that the error
in the determination of the electric current density increases very modestly as the spectral resolution degrades, with the notable exception
of $j_{\rm z}$, where $\sigma$ increases by as much as 5-10 \% from \emph{he1} to \emph{he8}.\\

\begin{table}
\caption{Powers of ten, $f$, that yield the {floor} values as $10^{-f}$, at two different optical depths 
($\tau_c=1$,$\tau_c=0.1$) for the different inversions carried out in Sect.~\ref{sec:noise_spec}.}
\label{tab:floor_noise_spec} 
\centering       
\begin{tabular}{cccccc}
\hline
 & {\bf he5} & {\bf he6} & {\bf he1} & {\bf he7} & {\bf he8} \\
\hline
$\|j_{\rm x}(\tau_c=1)\|$ & 1.99 & 1.97 & 1.85 & 1.85 & 1.85 \\
$\|j_{\rm y}(\tau_c=1)\|$ & 2.00 & 1.97 & 1.85 & 1.85 & 1.85 \\
$\|j_{\rm z}(\tau_c=1)\|$ & 1.51 & 1.48 & 1.36 & 1.41 & 1.42 \\
$\|\ve{j}(\tau_c=1)\|$ & 1.48 & 1.43 & 1.27 & 1.29 & 1.29 \\
\hline
$\|j_{\rm x}(\tau_c=0.1)\|$ & 2.12 & 2.12 & 2.08 & 2.09 & 2.07 \\
$\|j_{\rm y}(\tau_c=0.1)\|$ & 2.11 & 2.10 & 2.07 & 2.07 & 2.06 \\
$\|j_{\rm z}(\tau_c=0.1)\|$ & 1.57 & 1.54 & 1.41 & 1.44 & 1.45 \\
$\|\ve{j}(\tau_c=0.1)\|$ & 1.53 & 1.49 & 1.36 & 1.38 & 1.37 \\
\hline                    
\end{tabular}
\end{table}
\begin{table*}
\caption{Standard deviation in the inference of the electric current density, $\sigma$, at two different optical 
depths ($\tau_c=1,0.1$) for the different Stokes inversions described in Sect.~\ref{sec:noise_spec}.
\label{tab:sigma_perce_noise_spec}}
\centering       
\begin{tabular}{cccccc}
\hline
 $\|j_{\rm x}\|$ & {\bf he5} & {\bf he6} & {\bf he1} & {\bf he7} & {\bf he8} \\
\hline
$\sigma(\tau_c=1)$ & 0.85/0.70 & 0.85/0.70 & 0.85/0.70 & 0.86/0.71 & 0.87/0.72\\
$\sigma(\tau_c=0.1)$ & 0.90/0.78 & 0.90/0.78 & 0.90/0.79 & 0.91/0.80 & 0.91/0.80\\
\hline
\hline
 $\|j_{\rm y}\|$ \\
\hline
$\sigma(\tau_c=1)$ & 0.86/0.69 & 0.86/0.69 & 0.85/0.69 & 0.86/0.70 & 0.87/0.71\\ 
$\sigma(\tau_c=0.1)$ & 0.90/0.77 & 0.90/0.77 & 0.90/0.77 & 0.91/0.79 & 0.92/0.79\\ 
\hline
\hline
 $\|j_{\rm z}\|$ \\
\hline
$\sigma(\tau_c=1)$ & 0.95/0.86 & 0.95/0.86 & 0.95/0.86 & 0.97/0.89 & 1.02/1.01\\
$\sigma(\tau_c=0.1)$ & 1.00/0.82 & 1.00/0.82 & 1.03/0.82 & 1.04/0.84 & 1.08/0.93\\
\hline
\hline
 $\|\ve{j}\|$ \\
\hline
$\sigma(\tau_c=1)$ & 0.66/0.51 & 0.65/0.51 & 0.65/0.51 & 0.66/0.53 & 0.68/0.55\\ 
$\sigma(\tau_c=0.1)$ & 0.72/0.57 & 0.72/0.57 & 0.73/0.57 & 0.73/0.58 & 0.74/0.58\\ 
\hline
\hline
\end{tabular}
\end{table*}

\section{Conclusions}
\label{sec:conclu}

We have studied the accuracy to which electric currents in the lower solar photosphere can be inferred via the application of
Stokes inversion codes to the radiative transfer equation for polarized light. Increasingly
complex inversions were applied to a set of synthetic observations in multiple spectral lines obtained
from radiative three-dimensional MHS simulations. The Stokes inversions range from simple ME-like inversions 
to regularized $z$-dependent inversions that include MHS constraints.\\

While the simplest inversions are capable of retrieving the three components of the electric current density ($j_x$,$j_y$,$j_z$)
within a factor of 10 of the values present in the MHD simulations ($\sigma \approx 1.0$ dex), the most complex one
improves this to within a factor of 5 ($\sigma \approx 0.7$ dex ) of the values present in the MHD simulations.
Concerning the modulus $\|{\bf j}\|$ of the electric current density, the accuracy is $\sigma \approx 0.7$ dex for the ME-like 
inversion (factor of 5) and $\sigma \approx 0.5$ dex (factor of 3) for the MHS inversion. The improvement is not caused by one particular kind 
of inversion but rather by the cumulative effect of all new layers of complexity (number of spectral lines, number of nodes, 
using MHS instead of HE, etc.).\\

Our results suggest that the main sources of errors are the limitations of the inversion models themselves. This is also
supported by the fact that neither photon noise nor spectral resolution play an important role in the determination
of the electric current density.\\

The most complex inversion tested in this paper (Sect.~\ref{subsec:mhs}) allows  the modulus electric current
density to be determined in regions of relatively strong magnetic fields ($B > 300$~gauss) within a factor of 2 ($\sigma = 0.2$ dex)
of the currents present in the simulations. This might be accurate enough to allow us to determine regions
in the solar photosphere where Joule or Ohmic heating is occurring.\\

\begin{acknowledgements}
This work has received funding from the Deutsche Forschungsgemeinschaft (DFG project number 321818926) and from the 
European Research Council (ERC) under the European Union's Horizon 2020 research and innovation programme (SUNMAG, 
grant agreement 759548). JMB acknowledges travel support from the Spanish Ministry of Economy and Competitiveness 
(MINECO) under the 2015 Severo Ochoa Program MINECO SEV-2015-0548 and from the SOLARNET project that has received 
funding from the European Union’s Horizon 2020 research and innovation programme under grant agreement no 824135.
The Institute for Solar Physics is supported by a grant for research infrastructures of national importance from the 
Swedish Research Council (registration number 2017-00625). Special thanks to K.D.~Leka for suggestions that led to the
development of the work presented in this paper. This research has made use of NASA's Astrophysics Data System.
\end{acknowledgements}

\bibliographystyle{aa}
\bibliography{ms}

\end{document}